\documentclass[%
superscriptaddress, 
 amsmath, amssymb, 
 aps, 
 twocolumn, 
 prb, 
]{revtex4-2}

\usepackage{graphicx}
\usepackage{dcolumn}
\usepackage{bm}
\usepackage{color}
\usepackage{xr-hyper}
\usepackage{amsmath}
\usepackage{overpic}
\usepackage[colorlinks=true]{hyperref}

\usepackage{booktabs}

\begin{document}


\title{
Towards Identifying the PL6 Center in SiC: From First-Principles Screening to Hyperfine Validation of Competing Defect Candidates
}

\author{Xin Zhao}
\affiliation{Key Laboratory of Microscale Magnetic Resonance, School of Physical Sciences, University of Science and Technology of China, Hefei 230026, China}
\affiliation{CAS Center for Excellence in Quantum Information and Quantum Physics, University of Science and Technology of China, Hefei 230026, China}

\author{Mingzhe Liu}
\email{duguex@ustc.edu.cn}
\affiliation{Key Laboratory of Microscale Magnetic Resonance, School of Physical Sciences, University of Science and Technology of China, Hefei 230026, China}
\affiliation{CAS Center for Excellence in Quantum Information and Quantum Physics, University of Science and Technology of China, Hefei 230026, China}

\author{Yu Chen}
\affiliation{Key Laboratory of Microscale Magnetic Resonance, School of Physical Sciences, University of Science and Technology of China, Hefei 230026, China}

\author{Qi Zhang}
\affiliation{Key Laboratory of Microscale Magnetic Resonance, School of Physical Sciences, University of Science and Technology of China, Hefei 230026, China}
\affiliation{School of Biomedical Engineering and Suzhou Institute for Advanced Research, University of Science and Technology of China, Suzhou 215123, China}
\affiliation{Institute of Quantum Sensing and School of Physics, Zhejiang University, Hangzhou, 310027, China}

\author{Chang-Kui Duan}
\email{ckduan@ustc.edu.cn}
\affiliation{Key Laboratory of Microscale Magnetic Resonance, School of Physical Sciences, University of Science and Technology of China, Hefei 230026, China}
\affiliation{CAS Center for Excellence in Quantum Information and Quantum Physics, University of Science and Technology of China, Hefei 230026, China}
\affiliation{Hefei National Laboratory, University of Science and Technology of China, Hefei 230088, China}


\date{\today}

\begin{abstract}
  The PL6 color center in 4H-SiC, known for its excellent ambient-temperature spin and optical properties, has an unresolved microscopic origin. In this first-principles study, we systematically investigate potential structures to clarify its nature. We first rigorously examine the DV-antisite hypothesis (a divacancy paired with a carbon antisite, $\mathrm{C_{Si}}$), analyzing the energetic, electronic, and spin properties of various $V_\mathrm{Si}V_\mathrm{C}+\mathrm{C_{Si}}$ configurations. Two $\mathrm{C_{3v}}$-symmetric $\mathrm{kk+C_{Si}}$ complexes emerge as strong candidates within this framework. Subsequently, a critical comparison of hyperfine interaction signatures is performed between these candidates, the alternative OV model [specifically OV(hh) and OV(kk), an oxygen replacing C together with a Si vacancy], and experimental data. This analysis demonstrates that the OV(hh) structure more accurately reproduces PL6's hyperfine features. Furthermore, re-evaluation of the proposed OV(hk) model for the related PL5 center reveals zero-field splitting parameter $E$ inconsistencies with experimental results, suggesting that PL5 and PL6 may have distinct origins. These findings provide crucial theoretical insights and motivate targeted experimental validation for these quantum defects.
\end{abstract}

\maketitle

\section{\label{intro}Introduction}

Spin defects in solid-state systems are promising candidate systems for quantum information science\,\cite{spin_defect_review}. Among these candidates, divacancies in 4H-SiC have garnered significant attention due to their long coherence time\,\cite{sic_ms_coherence, sic_long_coherence}, spin-selective optical transitions\,\cite{spin-to-photon}, and telecom-wavelength fluorescence\,\cite{sic-telecom}. Photoluminescence (PL) centers, specifically PL1-8\,\cite{PL1-7, PL8}, have been reported in 4H-SiC. In particular, PL5 and PL6 have similar optical and spin characteristics to those of PL1--4\,\cite{ZFS_expt, PL1-7}, while exhibiting superior charge stability under optical excitation\,\cite{SiC_charge_control, PL6_stability}. Experiment\,\cite{ZFS_expt} and first-principles calculations\,\cite{SiC_zpl_test} have unambiguously identified PL1--4 as the hh, kk, hk, and kh divacancies ($V_\mathrm{Si}V_\mathrm{C}$), where the symbols refer to the hexagonal (h) or cubic (k) crystallographic sites of the vacancy pair. However, the  PL5-8 centers have not yet been conclusively identified. A schematic diagram illustrating the 4H-SiC stacking and representative vacancy complexes is shown in Fig.\,\ref{fig:sites}(c). 

PL5 and PL6 exhibit properties highly similar to the PL1--4 divacancy defects and are generally considered to be divacancies perturbed by their local environment. Initial hypotheses proposed that PL5 and PL6 correspond to divacancies (specifically kh and kk types) located near stacking faults\,\cite{VV-SF-model}. However, this model was challenged by subsequent experiments\,\cite{PL6_debate}, fueling ongoing debate and further investigation\,\cite{exp_recent_1,exp_recent_2}. Divacancies are typically generated via post-irradiation annealing\,\cite{vv_creation} of SiC samples with a multifaceted point defect landscape, including intrinsic vacancies, antisite defects, and potential residual impurities. Given this complex defect landscape, an alternative plausible hypothes emerges: PL5 and PL6 may be complexes formed by combining a divacancy paired with another simple point defect nearby. Two prominent recent proposals advance this line of thought. One identifies PL5 as a kh divacancy complexed with a nearby carbon antisite ($\mathrm{C_{Si}}$)\,\cite{PL5_new}, referred to here as the DV-antisite model. The other, known as the OV model, attributes both PL5 and PL6 to oxygen-related defect complexes ($\mathrm{O_C V_{Si}}$)\,\cite{PL6_other_identify}. Analogous to the divacancy notation, the OV configurations are differentiated by the hexagonal (h) or cubic (k) sites occupied by the $\mathrm{O_C}$ and $V_\mathrm{Si}$. Fig.\,\ref{fig:sites}(c) illustrates one such configuration: $\mathrm{O_C}V_\mathrm{Si}$(hk). Under the widely held assumption that PL5 and PL6 share a common origin, a critical question arises: which scenario, DV-antisite or OV, provides a more robust explanation for PL6?

In this work, we systematically investigate the DV-antisite scenario for PL6 to identify its most probable structural origins. Using first-principles screening, we first analyze $V_\mathrm{Si}V_\mathrm{C} + \mathrm{C_{Si}}/\mathrm{Si_C}$ configurations based on zero-field splitting (ZFS), Kohn-Sham (KS) gap (used as an indicator of optical transition energy), and binding energy. This screening identifies two $\mathrm{kk+  C_{Si}}$ structures as promising candidates for PL6 within the DV-antisite framework. To further assess their validity, we compare these candidates with the leading OV(hh) structure proposed in Ref.\,\cite{PL6_other_identify}, focusing on hyperfine interaction signatures. Our analysis suggests that the OV(hh) structure is the most probable origin of PL6. Additionally, we examine whether PL5 could be attributed to the proposed OV(hk) candidate, based on the hypothesis that PL5 and PL6 share a similar origin. However, inconsistencies between the calculated and measured ZFS $E$ parameter rule out the possibility that PL5 is a simple OV(hk) structure.

The remainder of this paper is organized as follows. Sec.\,\ref{sec:method} outlines our comprehensive first-principles computational methodology. This section details the general calculation setup (Sec.\,\ref{sec:method-comp}), our strategy for identifying and screening potential defect candidates for PL6 under the DV-antisite hypothesis (Sec.\,\ref{sec:field} and Sec.\,\ref{sec:screen}), and the approach used to calculate the hyperfine interactions for the subsequent validation (Sec.\,\ref{sec:hyper_calc}). Sec.\,\ref{sec:result} presents our findings. We first discuss the results of the screening within the DV-antisite framework to identify promising candidates in Sec.\,\ref{sec:result_dv}. This is followed by a critical validation step, comparing the hyperfine interaction signatures of DV-antisite candidates and alternative OV model structures against experimental data to determine the most probable origin of PL6 (Sec.\,\ref{sec:hyperfine}). We then extend our discussion to consider implications for the PL5 center (Sec.\,\ref{sec:pl5}). Finally, Sec.\,\ref{sec:conclusion} summarizes our conclusions.

\section{Methodology\label{sec:method}}

Our investigation into the origin of PL6 commences with a systematic evaluation of the DV-antisite hypothesis, where PL6 is modeled as a divacancy paired with a nearby antisite.
While the primary focus of the screening methodology detailed in this section is to identify promising candidates within this DV-antisite framework, the underlying first-principles calculation setup and characterization techniques are generally applicable and were consistently employed for all defect structures considered in this study, including the OV model. Thus, we begin by detailing the computational approaches for screening and characterizing these divacancy-antisite configurations.

\subsection{\label{sec:method-comp}Calculation setup}

The electronic structure calculations were performed using the Vienna {\it ab initio} Simulation Package (VASP)\,\cite{vasp-core1, vasp-core2}, employing the projector augmented wave (PAW) method\,\cite{PAW-method} to describe the core regions of the electronic wavefunctions with recommended PAW pseudopotentials adopted. The Revised Perdew-Burke-Ernzerhof for solids functional\,\cite{PBEsol} was utilized. The plane-wave cutoff energy was set to 520\,eV, with a 12$\times$12$\times$2 Monkhorst-Pack k-point mesh used for the primitive cell and a single $\Gamma$ point for the expanded supercell. The convergence criterion was set to $10^{-6}$\,eV for the self-consistent field calculation and 0.01\,eV/\AA\ for the ionic relaxation. The ZFS matrix was calculated using a formalism that employs KS orbitals from periodic density functional calculations\,\cite{ZFS_calc} with a home-built code. The PAW correction\,\cite{ZFS_PAW_corr} and spin decontamination\,\cite{ZFS_spin-decontamination} were not considered here, and the accounted error was considered in the final results instead. 

For the defect calculations, a cubic-like supercell was constructed to host the defect. When the defect was generated in the initial perfect cell, small random displacements were applied to all atomic coordinates, then the structure was relaxed until the force criterion was satisfied. This procedure was intended to exclude the influence of the initial configuration selection on the final obtained structure.

\subsection{\label{sec:field}Determining Candidate Complex Defects}

To identify candidates for PL6 within the DV-antisite framework, we first consider experimental constraints. Experimental observations show that the rhombic ZFS parameter $E$ of PL6 is, in essence,  zero within experimental uncertainties\,\cite{PL1-7, ZFS_expt}, indicating that the principal axis of PL6 aligns with the c-axis. Consistent with this, angle-resolved measurements also indicate a ${\rm C_{3v}}$-like symmetry for PL6\,\cite{PL6_debate}. Furthermore, crucial constraints arise from PL6's electronic and spin properties. PL6 exhibits an $S=1$ ground state. This $S=1$ state, similar to PL1--4, implies an even electron number and an electronic configuration resembling that of divacancies. Consequently, our search focuses on defect complexes that preserve these features. Introducing an additional vacancy is excluded as it would lead to $S \neq 1$. While neutral antisite defects act as isovalent substituents and maintain a comparable electronic environment when combined with a diavacancy, thereby preserving the fundamental spin and optical properties, highly charged states like $\pm 2$ are excluded due to their significant deviation from the divacancy's electronic structure. Therefore, this study exclusively considers neutral hh/kk-type defect complexes incorporating neutral antisite defects.

With these constraints in mind, we proceed to search for possible complex defect structures. To explore variants of hh and kk configurations, we construct a cubic-like supercell with basis vectors $5\bm{a}, 3\bm{a}+6\bm{b}, 2\bm{c}$, where $(\bm{a}, \bm{b}, \bm{c})$ are the primitive translation vectors of the primitive cell. In the first stage, a hh or kk defect was generated. Then, a cylinder was defined centered at this defect, with a height of 8\,\AA\ and a base radius of 7\,\AA, as the region to search for potential antisite locations. To reduce reduncancy, we leveraged the ${\rm C_{3v}}$ symmetry of the parent hh and kk divacancies to exclude symmetry-equivalent antisite locations within the cylinder. The final selection space consists of 116 symmetry-inequivalent antisite placements relative to the neutral divacancy: 56 for hh and 60 for kk. This broad configuration space is designed is capture all antisite placements with meaningful impact on the divacancy properties, considering the anisotropic spatial decay of spin density from the hh/kk center (see Fig.\,\ref{spin-density} in the Supplemental Material\,\cite{suppl}).

\subsection{\label{sec:screen}Screening Methodology for the PL6 Center Identification}

To systematically identify the most plausible microscopic origin of PL6 among the candidate divacancy-antisite complexes, we established a multi-criteria screening framework integrating spin, optical, and thermodynamic properties. The selection criteria were designed to balance computational feasibility with experimental relevance, emphasizing three  aspects: (1) consistency with experimentally measured ZFS parameters; (2) compatibility with photoluminescence characteristics through KS level analysis; (3) thermodynamic stability of defect complexes. This multi-property approach enables effective discrimination among thousands of possible defect configurations while maintaining manageable computational costs.

\subsubsection{ZFS Parameter Calibration and Error-Tolerant Thresholds}
The ZFS matrix $\mathbf{D}$ dominates the spin Hamiltonian of a defect center. With a proper choice of axes, it can be diagonalized, yielding two parameters that characterize the energy splitting of spin states\,\cite{D_and_E}:
\begin{equation}
  D = \frac{3}{2} D_{zz}, \quad E = \frac{1}{2} (D_{yy} - D_{xx}).
\end{equation}
Note that experimentally measured $E$ values correspond to their absolute magnitudes. The ZFS parameters ($D$, $E$) serve as critical fingerprints for spin defects. To address systematic errors in our DFT implementation, including PAW correction and spin decontamination errors, we performed benchmark calculations on the four well-characterized divacancies (PL1--4, i.e., hh, kk, hk, kh defects), as shown in Table\,\ref{tab:zfs}. The calculated $D$ values are systematically overestimated by about 10\%. For the $|E|$ values, our calculations exhibit a discrepancy of roughly $\pm 10\,\textrm{MHz}$ when compared to the experimental data. Notably, the $|E|$ value for a strictly hh or kk divacancy should be zero; however, a small nonzero value emerges due to the absence of symmetry constraints in spin-density calculations. To ensure comprehensive coverage of potential PL6 candidates while accounting for these errors, we adopted the following conservative thresholds:
\begin{itemize}
\item $D$ $\in$ [1.365, 1.706]\,GHz ($1.0–1.25$ times experimental $D$ = 1.365\,GHz)
\item $|E| \lesssim$ 20\,MHz (allowing a $\pm 20\,$MHz deviation from over the experimental value of $E = 0\,$MHz)
\end{itemize}
This error-tolerant criterion accommodates computational uncertainties and potential variations in local crystal environments.

\begin{table}[htbp]
  \caption{\label{tab:zfs} Experimental and calculated values of ZFS parameters  $D$ and $|E|$ for four divacancy configurations in SiC. Experimental data are from Ref.\,\cite{ZFS_expt}.}
  \begin{ruledtabular}
  \begin{tabular}{lcccc}
 Divacancies  & \multicolumn{2}{c}{$D$ (GHz)} & \multicolumn{2}{c}{$|E|$ (MHz)} \\
   & expt & calc & expt & calc \\
  \addlinespace
  PL1(hh) & 1.336 & 1.506 & 0 & 6.4 \\
  PL2(kk) & 1.305 & 1.455 & 0 & 4.3\\
  PL3(kh) & 1.222 & 1.395 & 82 & 61 \\
  PL4(hk) & 1.334 & 1.452 & 18.6 & 29
  \end{tabular}
  \end{ruledtabular}
\end{table}

\subsubsection{KS Level Difference as ZPL Predictor in Defect Screening}

The photoluminescence characteristics provide critical complementary signatures for defect identification. For the ${\rm C_{3v}}$ symmetry divacancies, the ZPL originates from the spin-allowed $^3\textrm{E} \rightarrow {}^3\textrm{A}_2$ transition\,\cite{SiC_electronic}, corresponding to two specific occupations of KS orbitals differed by promoting an electron from the highest occupied $a_1$ orbital to the lowest unoccupied $e$  orbital in the minority spin channel. The calculation of ZPL of PL1--4 has been detailed and tested among supercell size, k-point mesh, and functional selection\,\cite{SiC_zpl_test}. However, such calculations are computationally demanding (e.g., HSE06 functional, a $\Gamma$-only calculation with a supercell of 2400 atoms, or a calculation with $4\times 4\times 4$ $k$-mesh for a 96-atom supercell)\,\cite{SiC_zpl_test, HSE06}. 

The observed 0.1\,eV blueshift in ZPL of PL6 over the hh/kk divacancies\,\cite{ZFS_expt, SiC_charge_control} suggests a perturbation of the splitting between KS orbitals from an antisite defect. We establish a computationally efficient proxy for ZPL shifts through KS level analysis: the energy difference between the highest occupied ($a_1$) and lowest unoccupied ($e$) minority spin states systematically correlates with experimental ZPL trends in benchmarked defects (PL1--4). This relationship arises because ZPL energies and KS splitting respond proportionally to the perturbation. By requiring candidate complexes to exhibit enhanced KS splitting compared to isolated hh/kk centers, we effectively screen for structures capable of reproducing PL6's characteristic ZPL shift.

\subsubsection{Binding Energy for Stability of Complex Defects}
Given that PL6 likely forms through radiation-induced defect aggregation rather than thermal equilibrium processes, we employed binding energy ($E_b$) rather than formation energy to assess the relative stability of candidate defects. Taking hh as an example, the binding energy is defined as\,\cite{rmp_point_defects}:
\begin{equation}
  E_b = E^f[\textrm{hh} + \textrm{antisite}] - E^f[\textrm{hh}] - E^f[\textrm{antisite}], 
  \label{binding}
\end{equation}
where $E^f$ is the formation energy of point defects. Negative $E_b$ values indicate exothermic complex formation, serving as a necessary (though not sufficient) condition for stability. While full thermodynamical analysis would require considering formation pathways\,\cite{formation_pathway} and configurational entropy\,\cite{configure_entropy}, our screening conservatively requires $E_b < 0$ to filter energetically favorable complexes.

At this stage, by applying criteria derived from several key experimental observations for the PL6 center—namely, its ZFS parameters ($D = 1.365$\,GHz and $E = 0$\,MHz) and its ZPL blueshift relative to hh/kk divacancies—along with a theoretical requirement for thermodynamic stability, we aim to identify promising candidates. Specifically, the applied criteria are: D within $1.365 \times [1, 1.25]$\,GHz, $E$ within $[-20, 20]$\,MHz, KS gap exceeding the reference, and negative binding energy. These are applied to the generated divacancy-antisite configurations under the DV-antisite hypothesis. Despite the intentionally permissive nature of these initial criteria, the geometric and electronic constraints inherent to these specific complex defects yield only a few viable configurations, demonstrating remarkable inherent selectivity in defect screening.

\begin{figure*}[!htbp]
  \begin{minipage}[b]{0.58\textwidth}
    \begin{overpic}[width=\textwidth]{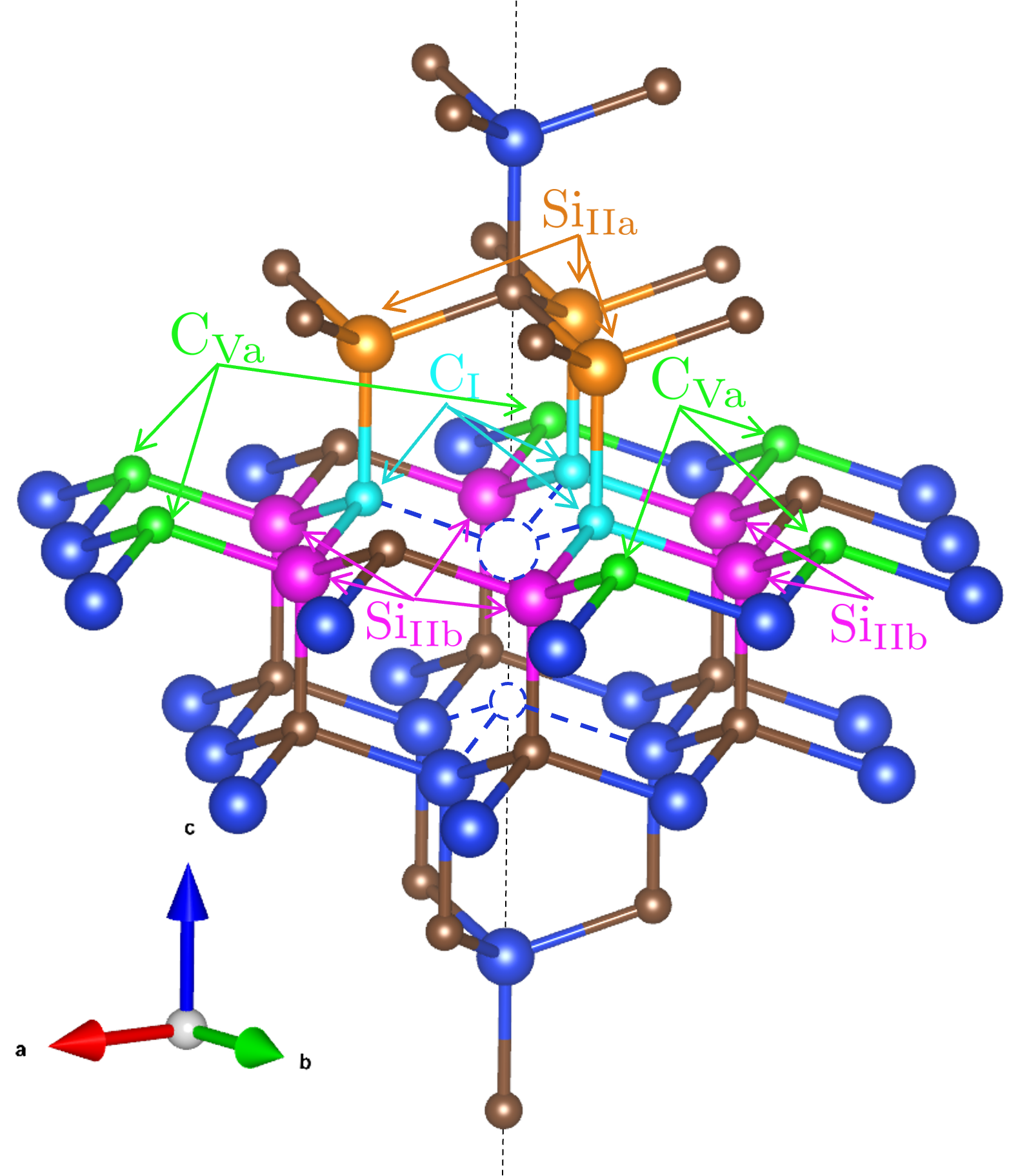}
      \put (5,90){\Large (a)}
    \end{overpic}
  \end{minipage}
  \hspace{0.02\textwidth}
  \begin{minipage}[b]{0.38\textwidth}
    \begin{overpic}[width=\textwidth]{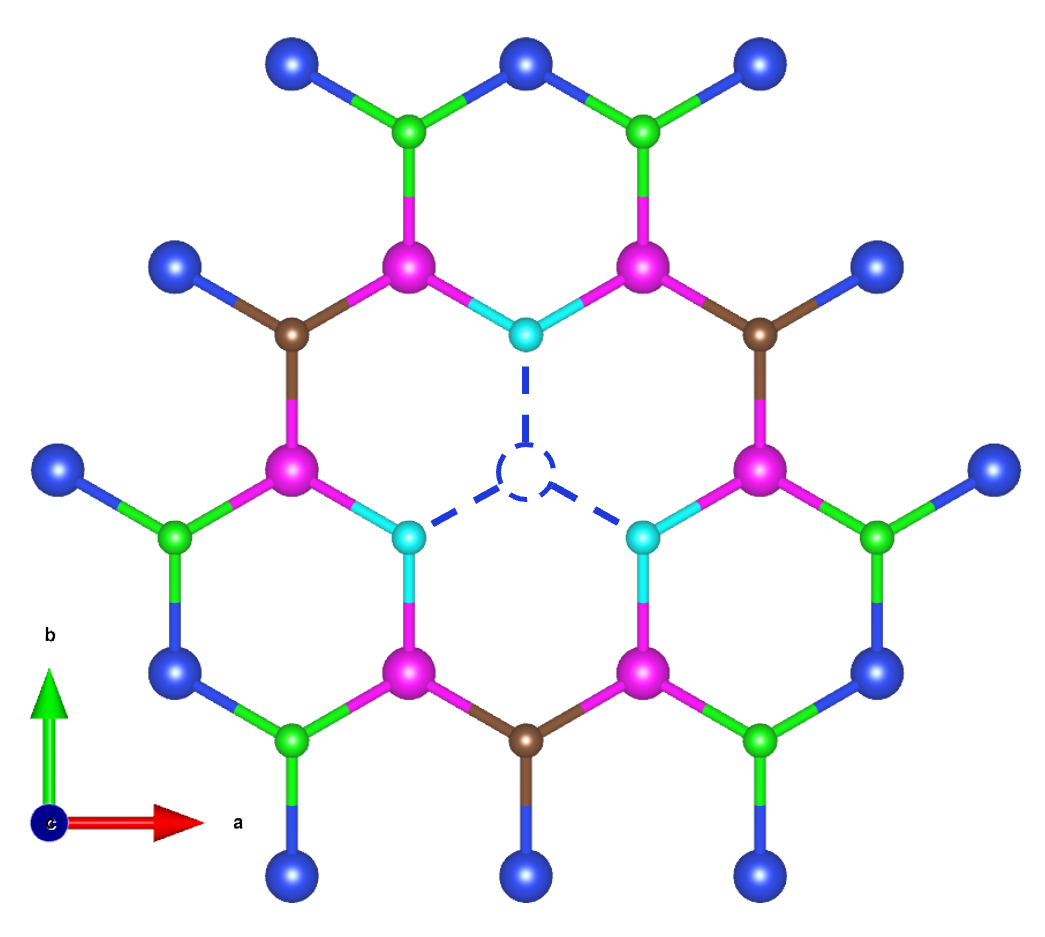}
      \put (5, 80){\Large (b)}
    \end{overpic}
    \vspace{2mm}
    \begin{overpic}[width=\textwidth]{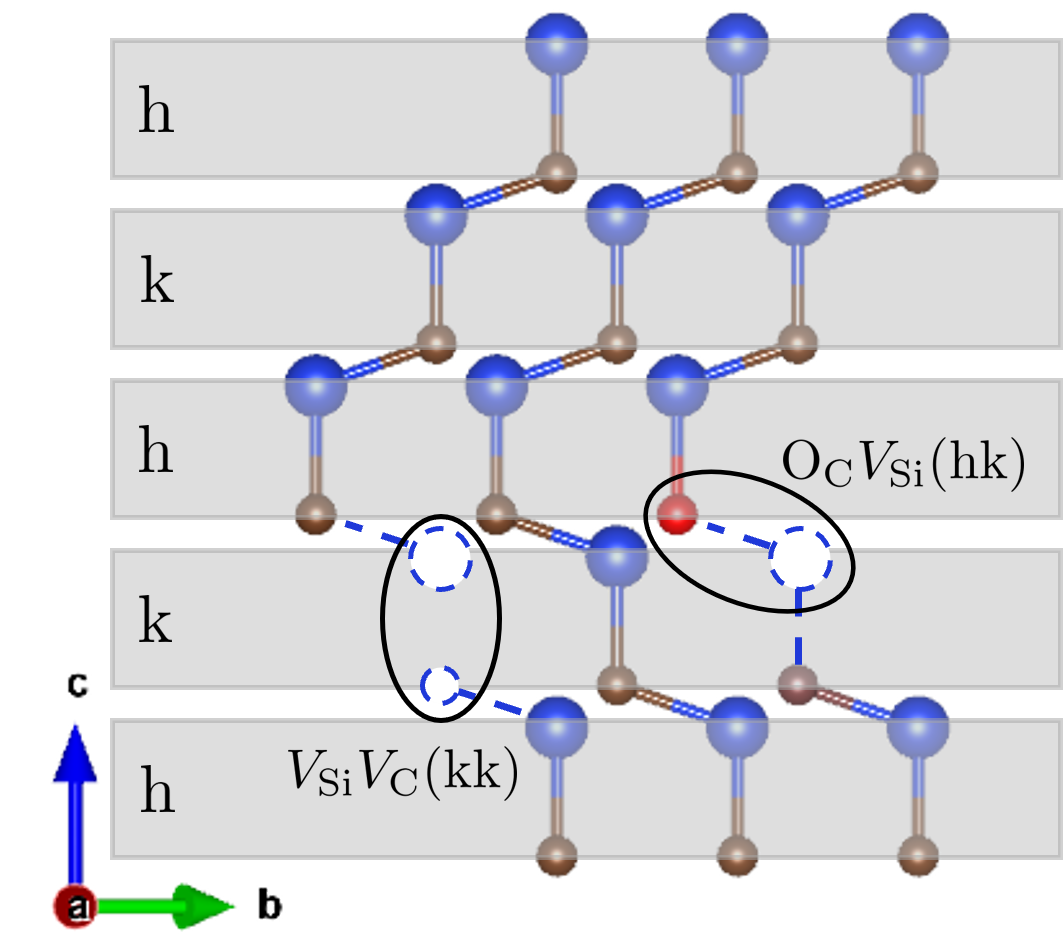}
      \put (0, 85){\Large (c)}
    \end{overpic}
  \end{minipage}
  \caption{\label{fig:sites}The structure, sites of 4H-SiC. (a) The local structure of a kk divacancy, illustrating important sites giving rise to hyperfine signatures typical for c-axis aligned defects in (a). (b) The top view of the plane containing the $V_\mathrm{Si}$, for better illustration of the labeled sites. In (a,b), the important sites are marked with different colors, and the label also adopts the same color. (c) Stacking order of atomic layers (h for hexagonal, k for cubic) along the c-axis in 4H-SiC, illustrating schematic examples of a $V_\mathrm{Si}V_\mathrm{C}$(kk) divacancy and an $\mathrm{O_C}V_\mathrm{Si}$(hk) defect.}
\end{figure*}

\subsection{\label{sec:hyper_calc}Hyperfine Interaction Calculation}
To further validate the candidates emerging from the screening process and to critically compare them against alternative models, we calculate their hyperfine interaction signatures. The hyperfine interation is described by a $3\times 3$ tensor ${\mathbf{A}^I}$ comprising isotropic part (Fermi contact, $\mathbf{A}_{\rm iso}^I$) and anisotropic ($\mathbf{A}_{\rm ani}^I$) components:

\begin{equation}
  (\mathbf{A}_{\text{iso}}^{I})_{ij} = \frac{2}{3} \frac{\mu_0 \gamma_e \gamma_I}{\langle S_z \rangle} \delta_{ij} \int \delta_T(\mathbf{r}) \rho_s (\mathbf{r} + \mathbf{R}_I) d\mathbf{r}, 
  \label{equ:Aiso}
\end{equation}

\begin{equation}
  (\mathbf{A}_{\text{ani}}^{I})_{ij} = \frac{\mu_0 \gamma_e \gamma_I}{4\pi \langle S_z \rangle} \int \frac{\rho_s(\mathbf{r} + \mathbf{R}_I)}{r^3} \frac{3 r_i r_j - \delta_{ij} r^2}{r^2} d\mathbf{r}, 
  \label{equ:Aani}
\end{equation}
where $\gamma_e$ and $\gamma_I$ are the gyromagnetic ratios of the electron and nuclear spin, respectively, $\rho_s$ is the spin density, $\delta_T$ is the Dirac delta function (approximated by a smeared form here), and $\mathbf{R}_I$ is the position of the nucleus\,\cite{hyperfine}. The total hyperfine tensor calculations were performed using the VASP code with the PAW method and the HSE06 functional.

To interpret these calculations in the context of experimental data and the specific nature of our candidate defects, several considerations are pertinent. For the DV-antisite candidates involving a $\mathrm{C_{Si}}$ antisite, the substitution of a Si atom with a C atom means that if this antisite nucleus significantly contributes to the hyperfine spectrum, the opposite sign of the $\mathrm{^{13}C}$ gyromagnetic ratio (compared to $\mathrm{^{29}Si}$) would lead to an inversion of the hyperfine coupling tensor's sign at that particular site. In the case of the OV models, while the $\mathrm{^{17}O}$ isotope possesses a nuclear spin, its extremely low natural abundance indicates that any associated hyperfine signals are too weak for detection in standard experiments without isotopic enrichment, making $\mathrm{^{29}Si}$ and $\mathrm{^{13}C}$ the primary contributors to observable hyperfine structure. While the sign of the hyperfine interaction can, in principle, be determined by more complex methods like electron-nuclear double resonance (ENDOR)\,\cite{ENDOR_hyper_sign}, the crucial experimental report that identified the hyperfine signatures of PL6 utilized ODMR, which resolves the magnitude of the splittings rather than the sign\,\cite{PL6_hyper}. Therefore, to perform the most direct and rigorous comparison with the available experimental findings, our analysis focuses on the principal component $A_{zz} = |\hat{n}_0 \cdot \mathbf{A}^I\cdot \hat{n}_0|$, where $\hat{n}_0$ is the unit vector along the c-axis, coinciding with the principal axis of the ZFS $\mathbf{D}$ tensor for the $\mathrm{C_{3v}}$-like centers.

\section{\label{sec:result}Results and Discussion}

\subsection{\label{sec:result_dv}Screening Potential PL6 Structures in DV-antisite Configurations}

Applying the multi-criteria screening protocol (detailed in Sec.\,\ref{sec:field}) to the set of generated divacancy-antisite defect configurations, we effectively narrowed the vast configuration space. This process isolated only 12 structures consistent with the primary experimental constraints for PL6, as listed in Table\,\ref{tab:results_sorted}. To systematically characterize these surviving configurations, we label the unperturbed divacancy and associated antisite defects using a grid-based positional encoding scheme. The numerical indices following an antisite symbol specify its position relative to the ${\rm V_{Si}}$ of the divacancy using grid coordinates along the $\bm{a}$-, $\bm{b}$-, and $\bm{c}$-axes, where negative displacements are denoted by overlined numbers. These coordinate axes are defined according to a gridded supercell with grid step sizes of 1.540\,\AA, 0.889\,\AA, and 0.628\,\AA\ along each respective axis. Complete implementation details, along with visual examples, are provided in Supplemental Material\,\cite{suppl}.

\begin{table}[htbp]
  \caption{\label{tab:results_sorted} Potential DV-antisite structures of PL6 and their calculated ZFS parameters $D$ and $E$, KS gap, and binding energy $E_b$. ZFS parameters of c-axis-oriented OV structures are also included. An en-dash (--) indicates not applicable.}
  \begin{ruledtabular}
    \begin{tabular}{lcccc}
      defect & $D$\,(GHz) & $E$\,(MHz) & KS gap\,(eV) & $E_b$\,(eV) \\ 
      \addlinespace
      hh & 1.506 & 6.4 & 1.117 & --\\
      kk & 1.455 & -4.3 & 1.141 & --\\
      \addlinespace
      OV(hh) & 1.594 & -7.2 & 1.181 & -- \\
      OV(kk) & 1.523 & 2.6 & 1.158 & -- \\
      \addlinespace
      $\mathrm{kk+  C_{Si}(0, 0, \overline{8})}$  & 1.478  & -4.3   & 1.176  & -0.204   \\
      $\mathrm{kk+  C_{Si}(0, 0, 8)}$  & 1.554  & 0.3   & 1.207  & -0.083   \\
      $\mathrm{kk+  C_{Si}(3, 3, 0)}$  & 1.477  & 18.7  & 1.151  & -0.035   \\
      $\mathrm{kk+  C_{Si}(\overline{1}, \overline{1}, \overline{12})}$  & 1.464  & 1.0   & 1.149  & -0.024   \\
      $\mathrm{kk+  C_{Si}(2, 0, \overline{8})}$  & 1.469  & -16.3  & 1.158  & -0.020   \\
      $\mathrm{hh+  C_{Si}(\overline{1}, \overline{1}, \overline{12})}$  & 1.518  & -10.0  & 1.127  & -0.014   \\
      $\mathrm{kk+  C_{Si}(0, \overline{2}, 12)}$  & 1.473  & 9.7   & 1.155  & -0.014   \\
      $\mathrm{hh+  C_{Si}(2, 2, \overline{12})}$  & 1.516  & -11.0  & 1.129  & -0.007   \\
      $\mathrm{hh+  C_{Si}(\overline{1}, \overline{1}, 12)}$  & 1.525  & -14.7  & 1.129  & -0.006   \\
      $\mathrm{hh+  C_{Si}(3, 1, \overline{8})}$  & 1.510  & 12.6  & 1.122  & -0.004   \\
      $\mathrm{kk+  C_{Si}(3, 1, 12)}$  & 1.461  & 13.7  & 1.142  & -0.004   \\
      $\mathrm{kk+  C_{Si}(3, \overline{3}, \overline{8})}$  & 1.459  & -13.3  & 1.142  & -0.001   \\
      \end{tabular}
  \end{ruledtabular}
\end{table}

A notable feature of the results is that most binding energies are small (on the order of $10^{-2}$\,eV or less). An exception is $\mathrm{kk+ C_{Si}(0, 0, \overline{8})}$, which has a more significant binding energy of -0.204\,eV. This reveals that for most candidates, the attraction between the divacancy and antisite defect is weak, leading to defect complexes with limited stability. Another feature is that most of the candidates possess an $E$ whose absolute value is in the range of 10 to 20\,MHz. Considering the precise experimental value of $E \approx 0$\,MHz for PL6, we further refine the selection by applying a more stringent of $|E| \leq 10\,$MHz to identify candidates best matching this key experimental signature. This yields the following candidates: $\mathrm{kk+ C_{Si}(0, 0, 8)}$, $\mathrm{kk+ C_{Si}(0, 0, \overline{8})}$, $\mathrm{kk+ C_{Si}(\overline{1},\overline{1},\overline{12})}$ and $\mathrm{kk+ C_{Si}(0,\overline{2},12)}$. Additionally, the remaining candidates show weak KS gap enhancement relative to the original kk divacancy defect. Among these candidates, $\mathrm{kk+ C_{Si}(0, 0, 8)}$ and $\mathrm{kk+ C_{Si}(0, 0, \overline{8})}$ exhibit the most enhanced KS level splitting. Considering all these aspects, $\mathrm{kk+ C_{Si}(0, 0, \overline{8})}$ and $\mathrm{kk+ C_{Si}(0, 0, 8)}$ emerge as the most compelling candidates for the PL6 center within the configuration space of DV-antisite complexes.

\begin{figure}[htbp]
  \centering
  \includegraphics[width=0.45\textwidth]{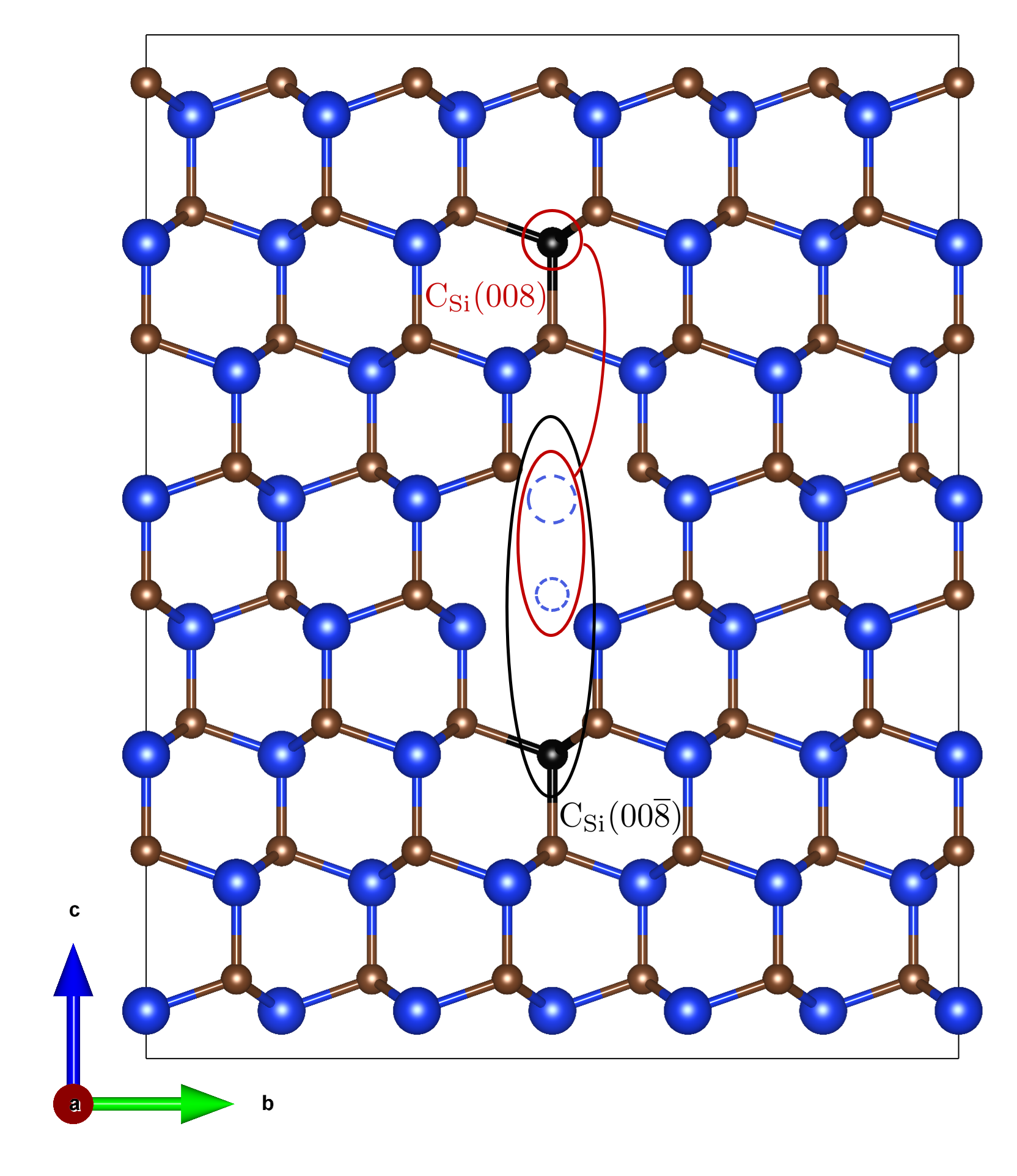}
  \caption{\label{fig:kk_most_possible}Atomic structure of the $\mathrm{kk+  C_{Si}(0, 0, \overline{8})}$ and $\mathrm{kk+  C_{Si}(0, 0, 8)}$ DV-antisite complexes. These configurations emerged as compelling candidates for the PL6 center from the screening of $V_\mathrm{Si}V_\mathrm{C}$+$\mathrm{C_{Si}}$ structures. The kk divacancy is shown as blue dotted circle, and the ${\rm C_{Si}}$ antisite defect is shown in black.}
\end{figure}

Fig.\,\ref{fig:kk_most_possible} shows the detailed atomic configurations of $\mathrm{kk+  C_{Si}(0, 0, \overline{8})}$ and $\mathrm{kk+  C_{Si}(0, 0, 8)}$ defect complexes. Both structures comprise a kk divacancy and a ${\rm C_{Si}}$ antisite defect along the $[001]$ direction. These configurations exhibit ${\rm C_{3v}}$ symmetry, consistent with experimental observations indicating that PL6 is a ${\rm C_{3v}}$-like center\,\cite{PL6_debate}. 

Apart from the difference in local environment between $\mathrm{kk+ C_{Si}(0, 0, \overline{8})}$ and $\mathrm{kk+  C_{Si}(0, 0, 8)}$, the dominant difference between them is the distance between the kk center and the antisite, which is 4.10\,\AA\ for $\mathrm{kk+  C_{Si}(0, 0, \overline{8})}$ and 6.00\,\AA\ for $\mathrm{kk+  C_{Si}(0, 0, 8)}$. The longer distance in $\mathrm{kk+  C_{Si}(0, 0, 8)}$ leads to a weaker interaction between the kk and the antisite, which is consistent with its smaller binding energy. For these two defects, we further calculated the actual ZPL energy by using constrained DFT calculation and the 0/1- charge transition level by the PBEsol functional. While this set of computational parameters cannot reliably reproduce experimental ZPL values, it should still capture the relative shifts between different configurations. This allows for a comparative analysis of trends despite the absolute discrepancy. A comparison of the results with kk is shown in Table\,\ref{tab:detail_info}. Notably, $\mathrm{kk+  C_{Si}(0, 0, 8)}$ exhibits a significant ZPL shift of 0.13\,eV relative to kk, whereas the ZPL shift for $\mathrm{kk+  C_{Si}(0, 0, \overline{8})}$ is considerably smaller, only 0.03\,eV. Moreover, the 0/-1 charge transition level of $\mathrm{kk+  C_{Si}(0, 0, 8)}$ is closer to the conduction band minimum (CBM) than kk, consistent with the experimental charge dynamics model\,\cite{SiC_charge_control}. Thus, although $\mathrm{kk+  C_{Si}(0, 0, 8)}$ has a relatively small binding energy, its ZPL shift and charge transition level characteristics establish it as a notable candidate for PL6 under the DV-antisite hypothesis. Although the ${\rm C_{Si}}$ is closer to the kk center in $\mathrm{kk+  C_{Si}(0, 0, \overline{8})}$, the properties of the divacancy mainly arise from the dangling bond of the C atom, which originates from the ${\rm V_{Si}}$ inside the divacancy. Although anti-site in $\mathrm{kk+  C_{Si}(0, 0, 8)}$ is farther from the kk center, it is closer to the dangling bond of the kk center, which is the reason why the impact of the antisite defect on the kk center is larger in $\mathrm{kk+  C_{Si}(0, 0, 8)}$ than in $\mathrm{kk+  C_{Si}(0, 0, \overline{8})}$.

To summarize,, we utilized experimental data to screen potential PL6 candidates, identifying the most likely structures under the DV-antisite scenario. However, there are also many other defects considered in our calculation that serve as good spin defects but are not PL6 candidates. They are listed in Table\,\ref{tab:other_spin_defects}. Some discussion on them is also given in the Supplemental Material\,\cite{suppl}.

\begin{table}[htbp]
  \caption{\label{tab:detail_info}The calculated ZPL and 0/-1 charge transition level of $\mathrm{kk+  C_{Si}(0, 0, \overline{8})}$ and $\mathrm{kk+  C_{Si}(0, 0, 8)}$, with kk also listed for comparison. $E^\mathrm{CBM}_{0/-1}$ represents the distance between the 0/1- level and the CBM.}
  \begin{ruledtabular}
    \begin{tabular}{lcc}
      defect & ZPL (eV) & $E^\mathrm{CBM}_{0/-1}$ \\
      \hline
      kk & 1.024 & 0.951 \\
      \hline
      $\mathrm{kk+  C_{Si}(0, 0, \overline{8})}$ & 1.055 & 0.962 \\
      $\mathrm{kk+  C_{Si}(0, 0, 8)}$ & 1.158 & 0.934
    \end{tabular}
  \end{ruledtabular}
\end{table}

\subsection{\label{sec:hyperfine}Further Validation Using the hyperfine parameter}

The previous screening identified $\mathrm{kk+  C_{Si}(0, 0, \overline{8})}$ and $\mathrm{kk+  C_{Si}(0, 0, 8)}$ as key candidates for PL6 within the DV-antisite framework. Concurrently, the c-axis-oriented OV(hh) and OV(kk) structures, previously proposed as potential divacancy-like centers\,\cite{PL6_other_identify}, also satisfy our initial screening criteria (as shown in Table\,\ref{tab:results_sorted}) and thus warrant rigorous comparison. To further refine the identification of PL6 and critically evaluate all these leading candidates, we now analyze their predicted hyperfine interaction signatures. As established in our methodology (Sec.\,\ref{sec:hyper_calc}), our analysis focuses on the principal component, $A_{zz}$, to allow for the most direct and rigorous comparison with available ODMR experimental data.

\begin{figure}
  \centering
  \includegraphics[width=0.45\textwidth]{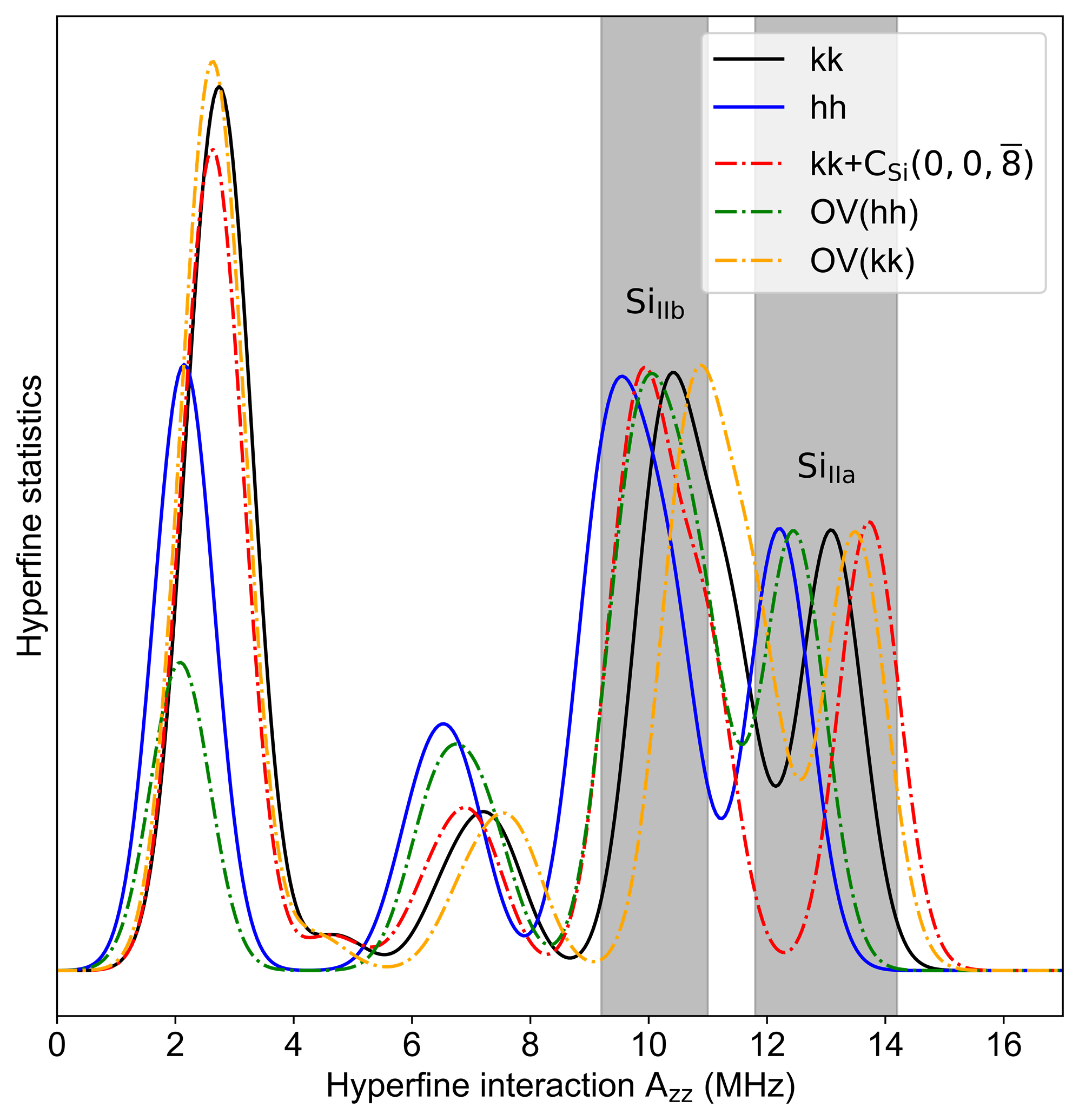}
  \caption{\label{hyper} The distribution of the $A_{zz}$ component for the kk, hh, $\mathrm{kk+  C_{Si}(0, 0, 8)}$ and OV(hh), OV(kk) defects. The smooth curve represents the sum of Gaussian distributions, with a mean value given by $A_{zz}$ and a standard deviation of 0.5\,MHz. The gray shaded regions correspond to the experimentally observed $\mathrm{Si_{IIa}}$ and $\mathrm{Si_{IIb}}$ hyperfine splitting sites. And the isotopic abundances of ${\rm ^{29}Si}$ and ${\rm ^{13}C}$ are also taken into account. For clearer individual presentations of these comparisions, Fig.\,\ref{hyper_stat}(b)-Fig.\,\ref{hyper_stat}(d) in the Supplemental Material\,\cite{suppl} each plot the $A_{zz}$ distribution of a specific key defect alongside those of hh and kk.}
\end{figure}

To initially visualize and compare the overall hyperfine landscape presented by different candidate structures, we examine the statistical distribution of these calculated $A_{zz}$ values for all relevant $\mathrm{^{29}Si}$ and $\mathrm{^{13}C}$ nuclei, as depicted in Fig.\,\ref{hyper}. This global representation offers an intuitive, qualitative assessment of similarities and differences in the density of hyperfine couplings across various frequency ranges. Fig.\,\ref{fig:sites}(a) and Fig.\,\ref{fig:sites}(b), which specifically depicts a kk divacancy, illustrates the typical atomic environment and key sites generally relevant for hyperfine interactions in various c-axis-oriented divacancies and related defects in 4H-SiC. The calculated $A_{zz}$ distributions in Fig.\,\ref{hyper} generally exhibit characteristic regions: weakly coupled distant nuclei ([1, 4]\,MHz), $\mathrm{^{13}C}$ at $\mathrm{C_{Va}}$ sites ([5, 8]\,MHz), and more strongly coupled $\mathrm{^{29}Si}$ at $\mathrm{Si_{IIb}}$ ([9, 12]\,MHz) and $\mathrm{Si_{IIa}}$ ([12, 15]\,MHz) sites\,\cite{DV_origin}, with a very strong signal from $\mathrm{^{13}C}$ at the $\mathrm{C_I}$ site around 80\,MHz (omitted for clarity). While these overall distributions provide a valuable qualitative first look, a more quantitative and decisive comparison with available experimental data necessitates focusing on specific, well-characterized hyperfine interactions. Therefore, our subsequent detailed analysis and comparison with experimental findings from Ref.\,\cite{PL6_hyper} will concentrate on the relative shifts of the well-characterized $\mathrm{Si_{IIb}}$ and $\mathrm{Si_{IIa}}$ sites.

Before proceeding to a detailed comparison of specific defect models, it is crucial to establish the basis of our validation against experimental hyperfine data. DFT calculations of hyperfine parameters can exhibit systematic deviations from experimental absolute values due to approximations in the exchange-correlation functional, pseudopotentials, and other computational factors\,\cite{calc_for_qubit}. However, relative shifts in hyperfine coupling strengths—such as the change observed when a nearby defect perturbs a divacancy, or differences between distinct defect configurations—are generally more robust and less susceptible to these systematic errors. The proven robustness of this comparative method makes it a reliable tool for validating the systems under investigation. Therefore, our primary validation strategy will focus on comparing these calculated relative shifts with experimentally observed trends, rather than on achieving precise agreement in absolute coupling constants. This approach allows for a more reliable assessment of which candidate structure best reproduces the experimentally characterized hyperfine signatures of PL6.

With this validation approach in mind, initial examination reveals that the $A_{zz}$ distribution for the $\mathrm{kk+C_{Si}(0, 0, \overline{8})}$ complex is nearly indistinguishable from that of the unperturbed kk divacancy (see Fig.\,\ref{hyper_stat} in the Supplemental Material\,\cite{suppl}(a)), making it an improbable source for the distinct hyperfine signature of PL6. More significantly, the $\mathrm{kk+ C_{Si}(0, 0, 8)}$ candidate, while initially promising based on other metrics, shows a notable discrepancy when its hyperfine predictions are compared to experiment. Specifically, our calculations for this complex predict a redshift of the $\mathrm{Si_{IIb}}$ peak (0.5\,MHz) but a blueshift of the $\mathrm{Si_{IIa}}$ peak (0.7\,MHz) relative to the kk divacancy. This pattern is inconsistent with the experimental findings for PL6, which report redshifts for both the $\mathrm{Si_{IIb}}$ (0.4\,MHz) and $\mathrm{Si_{IIa}}$ (0.7\,MHz) sites when compared to the kk reference\,\cite{PL6_hyper}.
Turning to the OV candidates, the OV(kk) complex can also be dismissed. Our calculations predict it would induce blueshifts at both sites(0.5\,MHz for $\mathrm{Si_{IIb}}$ and 0.4\,MHz for $\mathrm{Si_{IIa}}$) relative to the kk divacancy, again conflicting with the experimental shifts of PL6.
In striking contrast, the OV(hh) model exhibits excellent agreement with experimental trends. Our calculations for OV(hh) predict redshifts of 0.4\,MHz for $\mathrm{Si_{IIb}}$ and 0.6\,MHz for $\mathrm{Si_{IIa}}$ sites relative to the kk divacancy. Furthermore, when comparing to the hh divacancy as another reference, PL6 experimentally exhibits blueshifted $\mathrm{Si_{IIb}}$ and $\mathrm{Si_{IIa}}$ hyperfine couplings; the OV(hh) model successfully reproduces this, yielding calculated blueshifts of 0.5\,MHz($\mathrm{Si_{IIb}}$) and 0.2\,MHz ($\mathrm{Si_{IIa}}$). This consistent agreement in relative hyperfine shifts across multiple reference points and specific sites is compelling. Thus, the hyperfine analysis, grounded in the comparison of these robust relative changes, provides strong evidence identifying OV(hh) as the PL6 center.

\subsection{\label{sec:pl5}Discussion on the PL5 center}

Our preceding analysis, particularly the hyperfine validation comparing potential DV-antisite structures against OV models, strongly indicates that OV(hh) is the PL6 center. An OV(hk) origin for the PL5 center was proposed in Ref.\,\cite{PL6_other_identify}, based on its calculated ZFS $D$ value, ZPL, and formation energy. However, for a basal defect like hk-$V_\mathrm{Si}V_\mathrm{C}$ (and by extension, OV(hk)), the nonzero ZFS $E$ value is also a critical parameter for identification, which has not been explicitly discussed for OV(hk) previously. We therefore calculated the ZFS $E$ value for OV(hk) and found its magnitude to be 96.9\,MHz. Similarly, an $E$ value of 150\,MHz has been reported by another independent calculation\,\cite{ov_calculation}. Compared to the experimental $|E|$ value for PL5, which is approximately only 16\,MHz\,\cite{PL1-7,PL6_E_2}, the significantly larger calculated $E$ values (96.9\,MHz and 150\,MHz) represent a pronounced discrepancy that is unlikely to stem from computational inaccuracies alone. Therefore, our findings do not support the OV(hk) configuration as the origin of the PL5 center.

These findings have notable implications, as PL5 and PL6 are often assumed to share a common origin due to their closely matched  ZPL energies\,\cite{PL1-7,PL6_E_2,ZFS_expt}, stability in charge states\,\cite{SiC_charge_control}, and qualitatively similar relative concentrations\,\cite{PL5_new} observed experimentally. However, our results indicate a divergence: while OV(hh) emerges as the most probable candidate for PL6, the PL5 center cannot be straightforwardly attributed to a simple OV(hk) structure. However, it remains possible that PL5 originates from a more intricate configuration beyond the $\mathrm{kh+C_{Si}}$ model 
proposed in Ref.\,\cite{PL5_new}. 

The potential distinction in origin between PL5 and PL6 can be considered in at least two ways:
\begin{itemize}
  \item Both PL5 and PL6 are fundamentally related to oxygen-vacancy, but PL5 involves a more complex structural variation than the simple OV(hk), thereby maintaining a `common family' association.
  \item PL5 and PL6 arise from fundamentally different sources: PL6 is OV(hh)-based, whereas PL5 may originate from an entirely different defect type, such as a divacancy-antisite complex ($V_\mathrm{Si}V_\mathrm{C} + \mathrm{C_{Si}}$).
\end{itemize}
Distinguishing between these two scenarios is essential. Future experimental investigations, including controlled oxygen doping and subsequent analysis of PL5 and PL6 concentrations and properties, could yield valuable insights.  Additionally, theoretical studies should extend the search to encompass more intricate oxygen-related defect configurations beyond simple $\mathrm{O_C V_{Si}}$ structures for PL5, complemented by experiment explorations.

\section{\label{sec:conclusion}Conclusion}

In conclusion, we have performed a systematic first-principles investigation towards elucidating the microscopic origin of the PL6 center in 4H-SiC. Our study initially examined the DV-antisite hypothesis, screening divacancy-antisite ($V_\mathrm{Si}V_\mathrm{{C}}+\mathrm{C_{Si}}/\mathrm{Si_C}$) configurations based on zero-field splitting (ZFS), Kohn-Sham (KS) gap, and binding energy. This analysis identified two promising $\mathrm{kk+ C_{Si}}$ structures, specifically $\mathrm{kk+ C_{Si}(0, 0, \overline{8})}$ and $\mathrm{kk+ C_{Si}(0, 0, 8)}$, as leading candidates within this framework, both exhibiting the requisite $\mathrm{C_{3v}}$ symmetry consistent with experimental observations.

To further validate these candidates, we performed a critical comparison of their predicted hyperfine interaction signatures of the most promising DV-antisite candidates with experimental data for PL6 and with those calculated for the alternative OV models. This comparative analysis demonstrated that the OV(hh) structure accurately reproduces the experimental hyperfine signatures, identifying it as the origin of PL6.
Additionally,  our re-evaluation of the proposed OV(hk) structure for PL5 identified significant discrepancies between its calculated ZFS $E$ parameter and experimental observations, raising doubt about its viability as the PL5 center. This indicates that PL5 and PL6 have distinct microscopic origins.

These findings underscore the necessity of integrating multiple theoretical metrics with diverse experimental signatures to achieve reliable  defect identification. Future targeted ESR/ODMR experiments, specifically designed to probe the distinct hyperfine signatures of OV(hh) versus $kk+\mathrm{C_{Si}}$ complexes, will be crucial for definitively confirming the PL6 structure. Furthermore, our broader screening has identified additional divacancy-antisite complexes that, while not corresponding to PL6, may represent other stable spin defects in SiC worthy of future investigation.
\\
\begin{acknowledgments}
This work was supported by Innovation Program for Quantum Science and Technology (Grant Nos.
2021ZD0302200 and 2021ZD0303204), the National Natural Science Foundation of China (Grant Nos.\ 12474242 and 12274396), the Anhui Initiative in Quantum Information Technologies (Grant No.\ AHY050000), the CAS Project for Young Scientists in Basic Research, the Fundamental Research Funds for the Central Universities, and the China Postdoctoral Science Foundation (Grant Nos.\ 2021M703110 and 2022T150631). 
  The numerical calculations in this paper have been done on the supercomputing system in the Supercomputing Center of University of Science and Technology of China. The vaspkit code\,\cite{VASPKIT} was used to generate calculation files and analyze the results. Vesta\,\cite{Vesta} was used to visualize the structure. 
\end{acknowledgments}

\bibliographystyle{apsrev4-2}
\bibliography{main-format-re}

\end{document}



\title{Supplemental Material for \\
Towards identifying the PL6 Center in SiC: From First-Principles Screening to Hyperfine Validation of Competing Defect Candidates}

\author{Xin Zhao}
\affiliation{Key Laboratory of Microscale Magnetic Resonance, School of Physical Sciences, University of Science and Technology of China, Hefei 230026, China}
\affiliation{CAS Center for Excellence in Quantum Information and Quantum Physics, University of Science and Technology of China, Hefei 230026, China}

\author{Mingzhe Liu}
\email{duguex@ustc.edu.cn}
\affiliation{Key Laboratory of Microscale Magnetic Resonance, School of Physical Sciences, University of Science and Technology of China, Hefei 230026, China}
\affiliation{CAS Center for Excellence in Quantum Information and Quantum Physics, University of Science and Technology of China, Hefei 230026, China}

\author{Yu Chen}
\affiliation{Key Laboratory of Microscale Magnetic Resonance, School of Physical Sciences, University of Science and Technology of China, Hefei 230026, China}

\author{Qi Zhang}
\affiliation{Key Laboratory of Microscale Magnetic Resonance, School of Physical Sciences, University of Science and Technology of China, Hefei 230026, China}
\affiliation{School of Biomedical Engineering and Suzhou Institute for Advanced Research, University of Science and Technology of China, Suzhou 215123, China}
\affiliation{Institute of Quantum Sensing and School of Physics, Zhejiang University, Hangzhou, 310027, China}

\author{Chang-Kui Duan}
\email{ckduan@ustc.edu.cn}
\affiliation{Key Laboratory of Microscale Magnetic Resonance, School of Physical Sciences, University of Science and Technology of China, Hefei 230026, China}
\affiliation{CAS Center for Excellence in Quantum Information and Quantum Physics, University of Science and Technology of China, Hefei 230026, China}
\affiliation{Hefei National Laboratory, University of Science and Technology of China, Hefei 230088, China}

\date{\today}

\maketitle
\clearpage

\section{Spin density distribution of kk structure}

\begin{figure*}[htbp]
    \centering
    \begin{overpic}[width=0.495\textwidth]{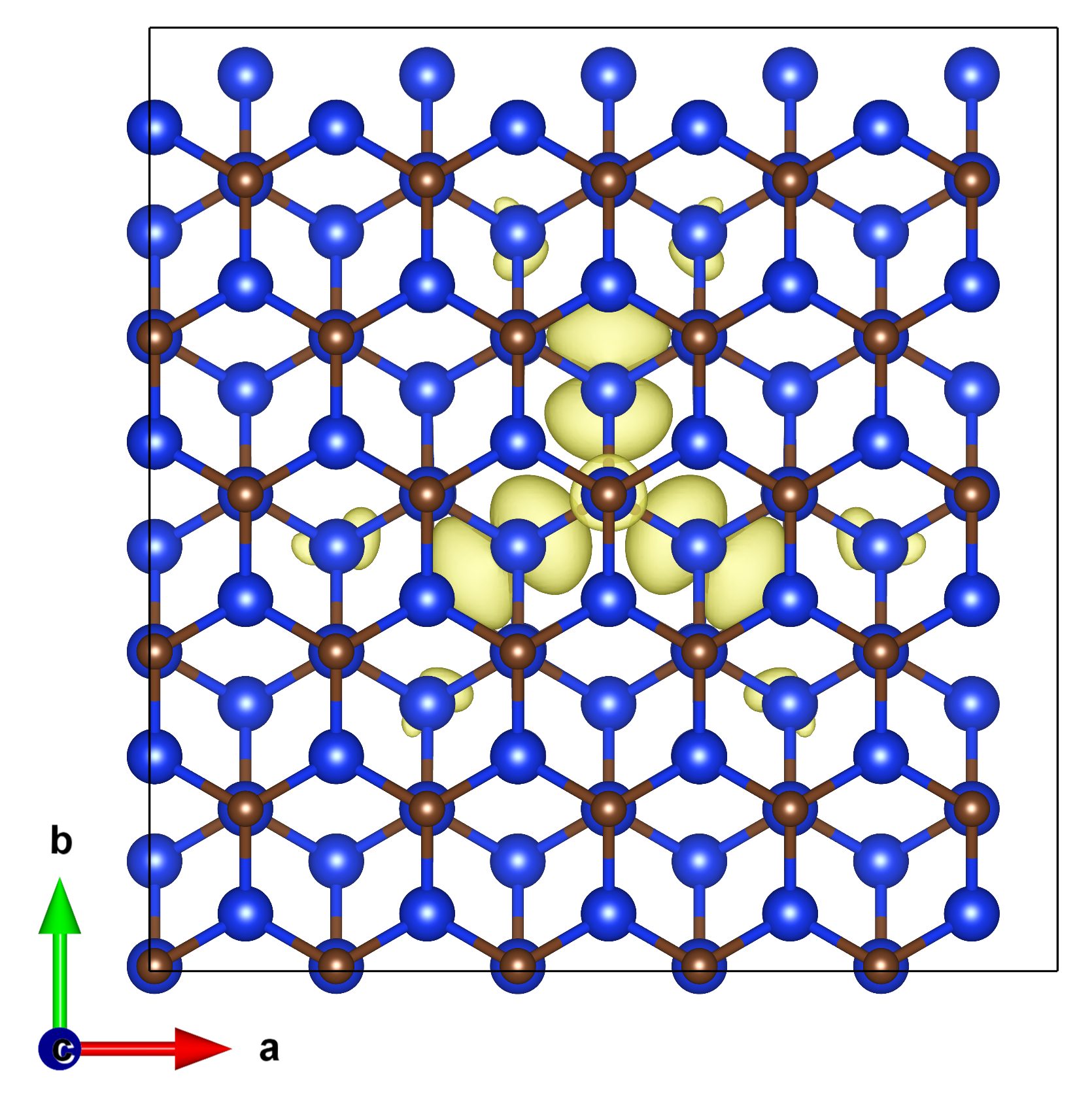}
      \put (0, 100) {\large (a)}
    \end{overpic}
    \begin{overpic}[width=0.465\textwidth]{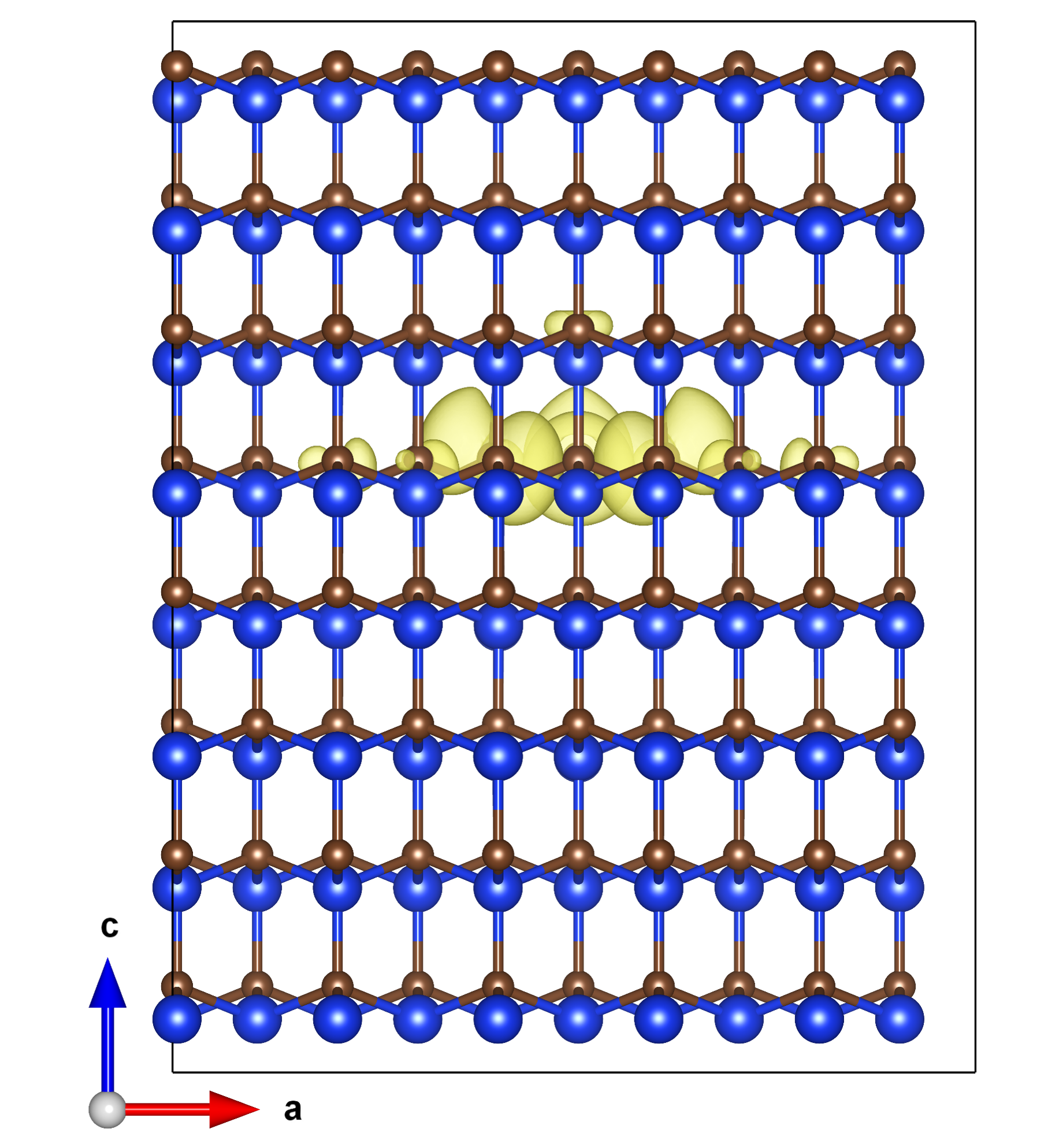}
      \put (0, 100) {\large (b)}
    \end{overpic}
    \caption{\label{spin-density}Spin density distribution of the kk structure. (a) Top view; (b) Side view.}
  \end{figure*}

Fig.~\ref{spin-density} shows the spin density distribution of the kk structure. The overall distribution respects the $\mathrm{ C_{3v}}$ symmetry of the kk center. The spin density is mainly distributed around the $V_\mathrm{ {Si}}$ center, which generates dangling C bonds. However, the spin density does not decay uniformly with distance from the $V_\mathrm{ {Si}}$ center. The distribution is quite extended and directional within the ab plane. The zero-field splitting (ZFS) parameters and hyperfine interaction rely heavily on the spin density distribution. Therefore, when considering possible antisite locations that could perturb the kk divacancy to form divacancy-antisite (DV-antisite) complexes, one should not only consider the nearest neighbor sites but also other sites that are far from the center, as they may also possess relatively high spin density and should be taken into account. For the sake of generality, we consider a relatively large selection range for the possible sites of the PL6 center, which is shown in the main text.
\newpage
\section{DV-antisite Nomenclature}
\begin{figure}[!htbp]
  \centering
  \begin{overpic}[width=0.495\textwidth]{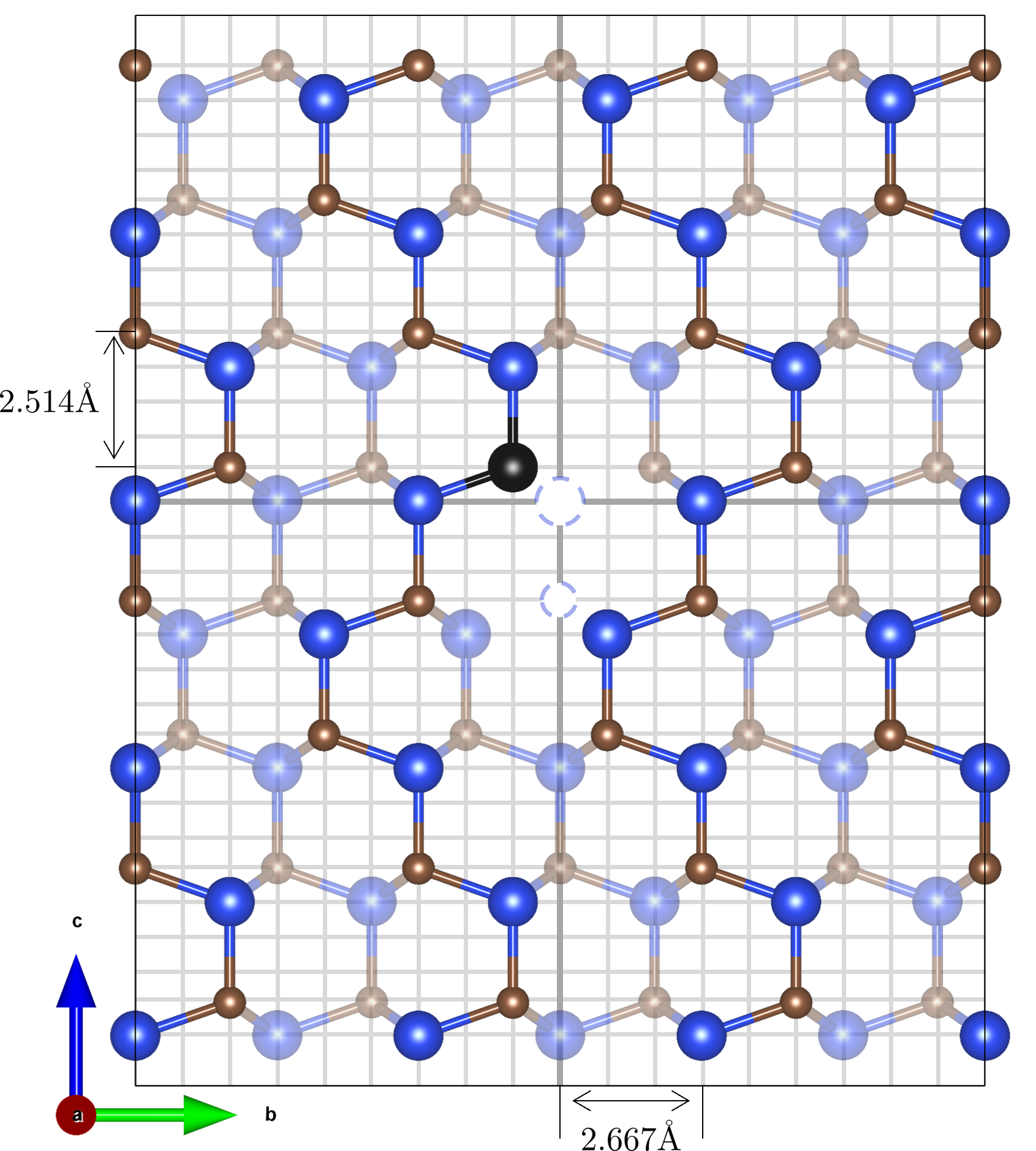}
    \put (4,95) {\large (a)}
  \end{overpic}
  \begin{overpic}[width=0.465\textwidth]{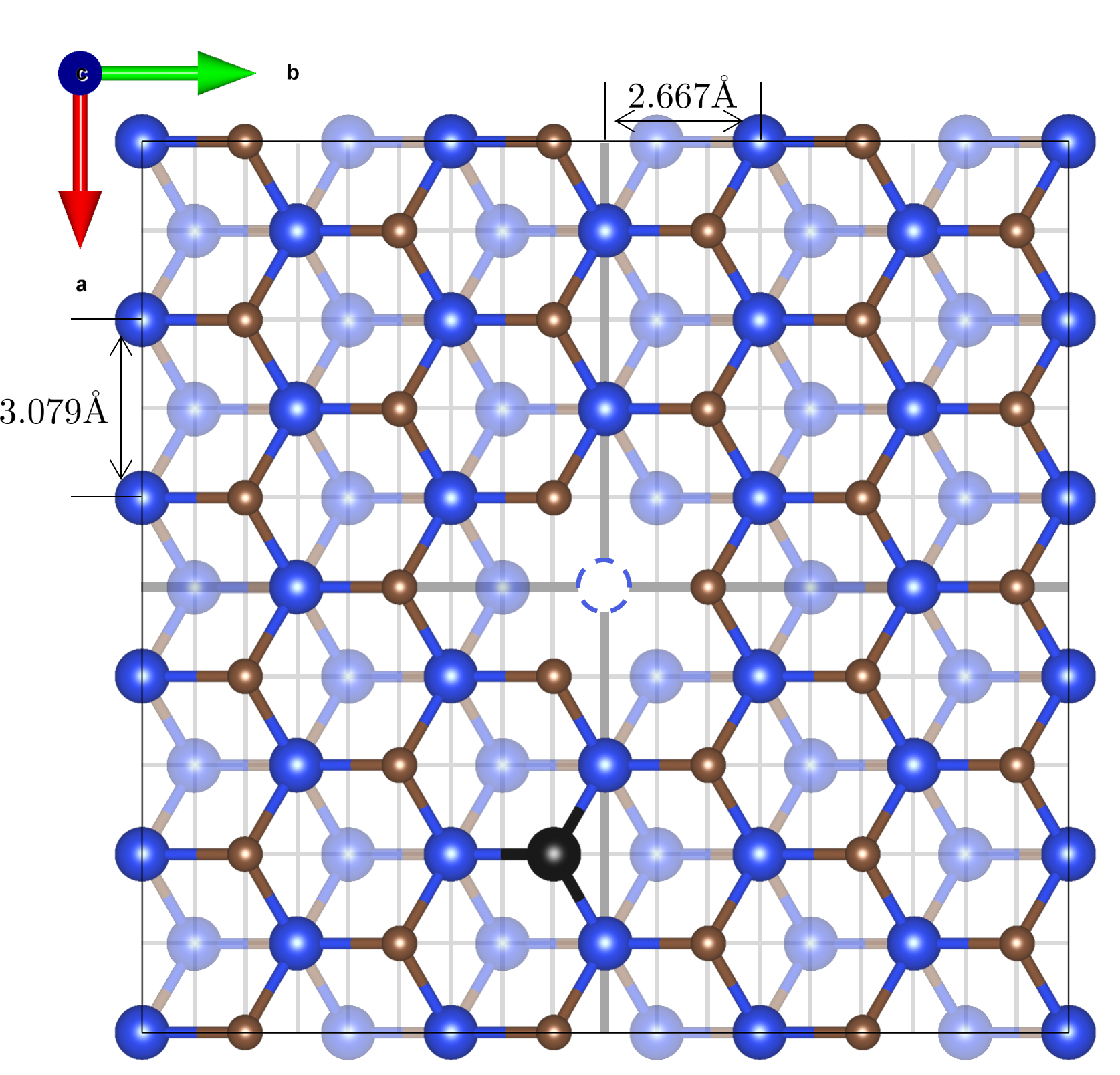}
    \put (4,95) {\large (b)}
  \end{overpic}
  \caption{\label{naming}Illustration of the grid division and nomenclature, using $\mathrm{kk+ Si_C}(3,\overline{1},1)$ as an example. The blue dashed circle indicates the vacancy position, while the dark atom represents the $\mathrm{ C_{Si}}$ antisite. The $V_\mathrm{ {Si}}$ location (coordinate origin) is marked  with thicker, darker-colored grid lines. Arrows in (a) and (b) denote the positive directions of the $\bm{a}$-, $\bm{b}$-, and $\bm{c}$-axes. Actual grid distances are provided in the figure. (a) Side view showing grids spanning vectors $\bm{b}$ and $\bm{c}$. (b) Top view showing grids spanning vectors $\bm{a}$ and $\bm{b}$.}
\end{figure}

To describe the DV-antisite structure, we propose a naming convention: \textit{divacancy} + \textit{antisite}(\textit{relative coordinates}). Here, the \textit{divacancy} refers to a predefined divacancy configuration, and the relative coordinates indicate the antisite defect's position relative to the $V_\mathrm{ {Si}}$ in the divacancy. The supercell is partitioned into uniform grids along lattice vectors $\bm{a},\bm{b}$, and $\bm{c}$, with spacings of 1.540~Å, 0.889~Å, and 0.628~Å, respectively. This ensures that all atomic sites correspond to grid points, and relative coordinates can be expressed as integer grid indices along these vectors, with the site of $V_\mathrm{ {Si}}$ serving as the origin $(0,0,0)$. Negative coordinates are denoted by overbars (e.g., $-1$ becomes $\overline{1}$). For example, in $\mathrm{kk+ Si_C}(3,\overline{1},1)$, the $\mathrm{ Si_C}$ antisite is located 3 grid units along $\bm{a}$, $-1$ unit along $\bm{b}$, and 1 unit along $\bm{c}$ relative to the $V_\mathrm{ {Si}}$ origin, as illustrated in Fig.~\ref{naming}.
\newpage

\section{Hyperfine Statistics of the possible PL6 center}
Identifying the microscopic origin of PL6 requires distinguishing between several leading candidate defect structures, including DV-antisite complexes and OV defects. The introduction of a nearby defect, such as a $\mathrm{C_{Si}}$ antisite or an $\mathrm{O_C}$ substituting atom, perturbs the electronic structure and spin density distribution of the parent divacancy. This perturbation leads to changes in the hyperfine interactions with nearby nuclear spins. 
While unique hyperfine signatures might be expected in principle – such as a sign reversal at the $\mathrm{C_{Si}}$ site itself (due to the opposite Si/C gyromagnetic ratios) or a contribution from $\mathrm{^{17}O}$ in the OV case - these effects are generally inaccessible in standard ESR/ODMR experiments. The sign reversal is not directly measurable via fine structure splitting, and the low abundance of $\mathrm{^{17}O}$ makes its signal undetectable without isotopic enrichment.
Therefore, to bridge theory and experiment and validate candidate structures against the experimental hyperfine signatures of PL6, we focus on the relative shifts in the hyperfine coupling parameters of well-characterized nuclei, specifically the $\mathrm{Si_{IIa}}$ and $\mathrm{Si_{IIb}}$ sites, compared to reference defects (kk and hh divacancies). While absolute calculated hyperfine values from DFT can have systematic errors, these relative shifts are generally more robust and provide a critical fingerprint for defect identification.

\begin{figure}[!htbp]
    \centering
    \begin{overpic}[width=0.48\textwidth]{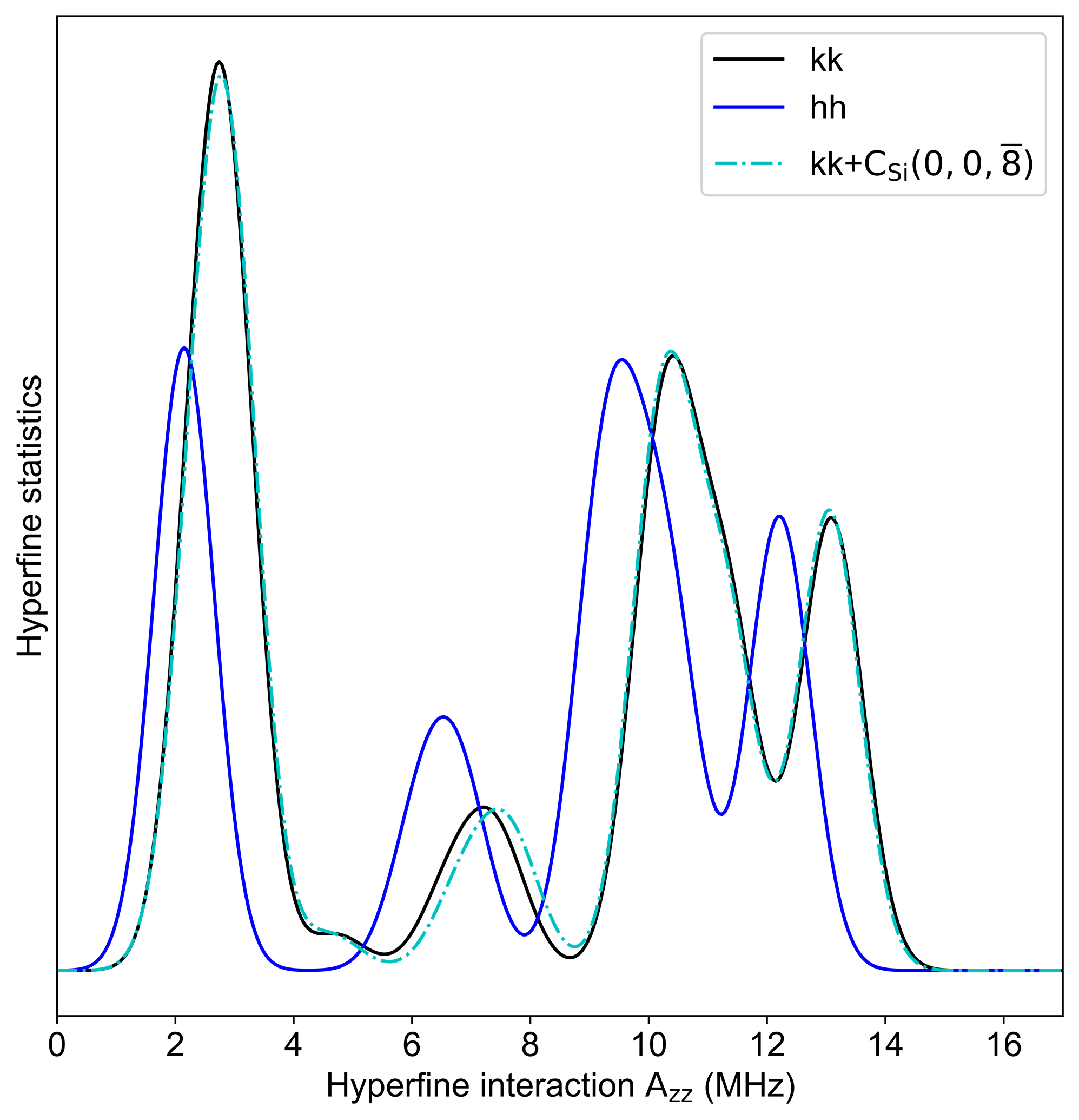}
      \put (5,91) {\large (a)}
    \end{overpic}
    \begin{overpic}[width=0.48\textwidth]{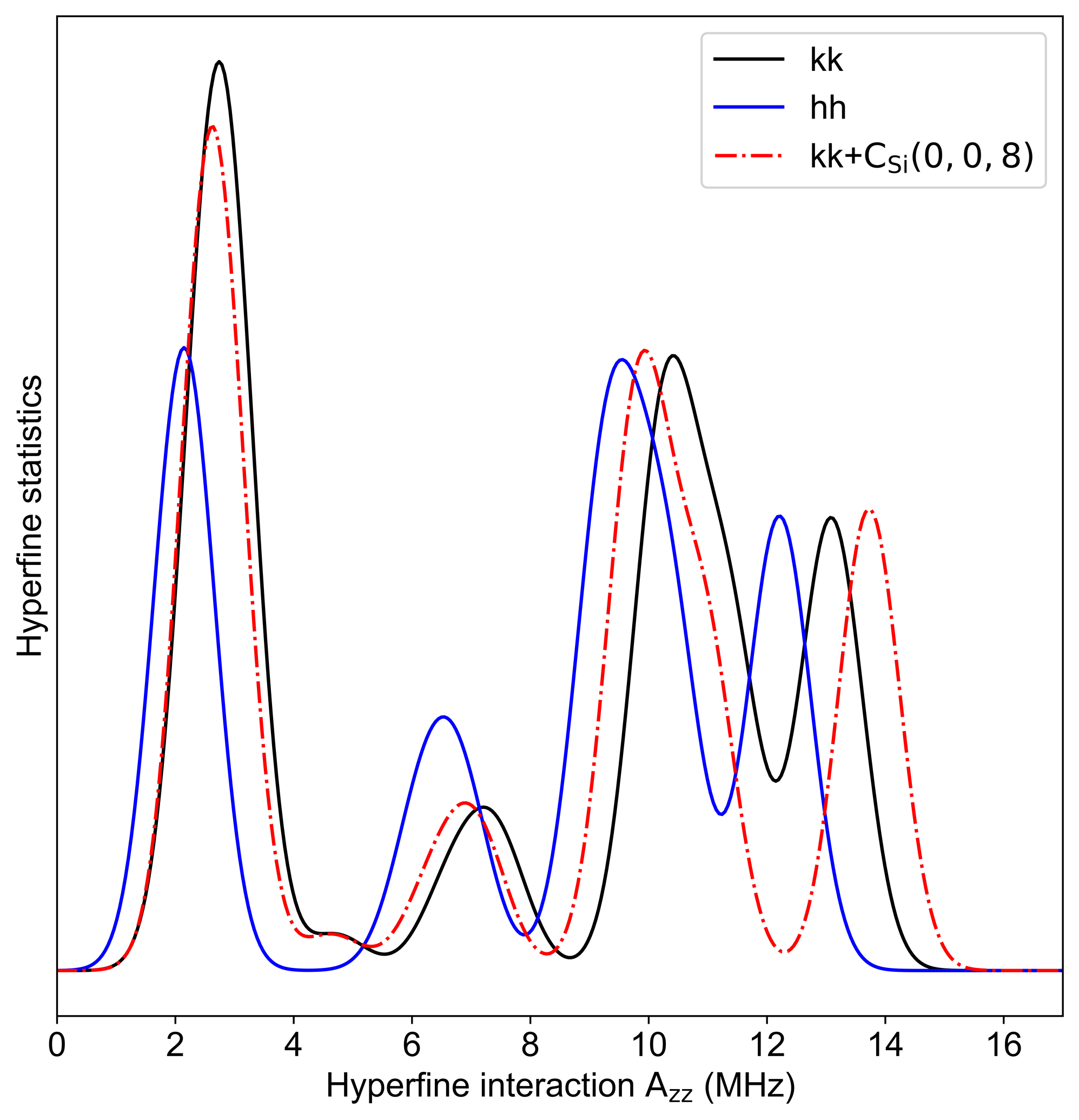}
      \put (5,91) {\large (b)}
    \end{overpic} \\
    \begin{overpic}[width=0.48\textwidth]{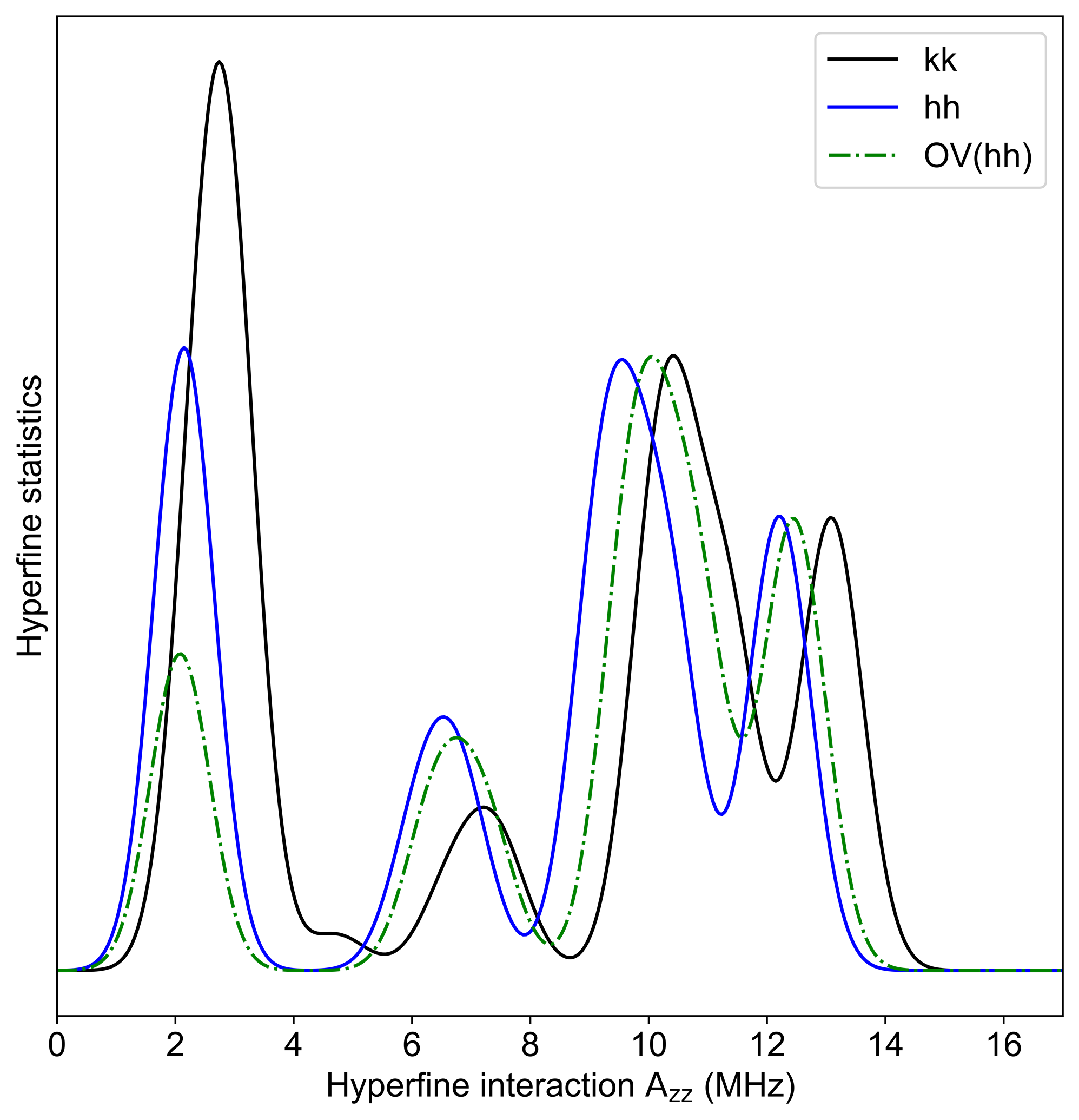}
      \put (5,91) {\large (c)}
    \end{overpic}
    \begin{overpic}[width=0.48\textwidth]{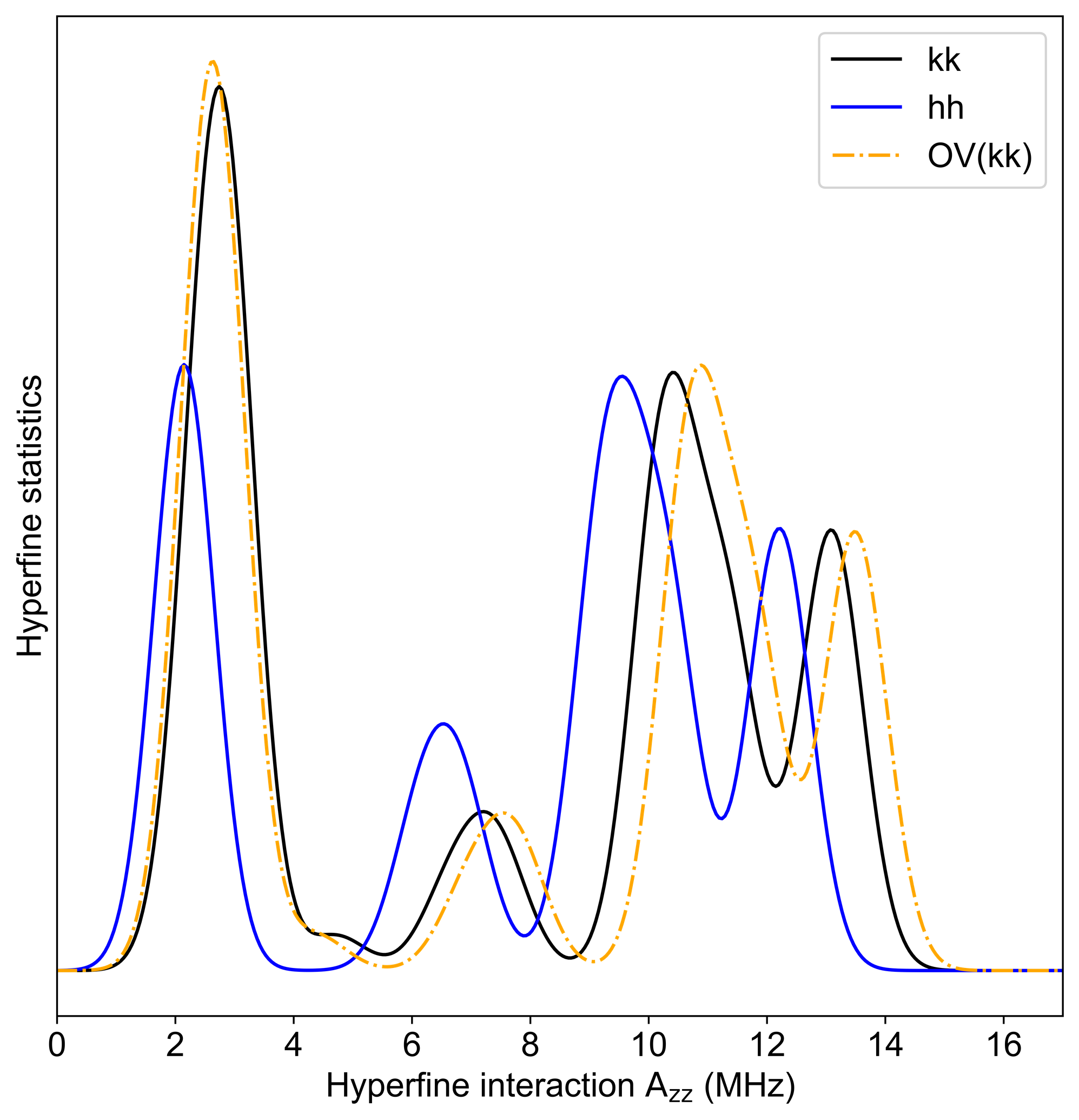}
      \put (5,91) {\large (d)}
    \end{overpic}
    \caption{\label{hyper_stat}Hyperfine statistics of two $\mathrm{kk + C_{Si}}$ candidate complexes and two c-axis-oriented OV defects compared to the kk and hh divacancy. (a) The $\mathrm{kk+ C_{Si}(0,0,8)}$ configuration; (b) The $\mathrm{kk+ C_{Si}(0,0,\overline{8})}$ configuration; (c) The OV(hh) configuration; (d) The OV(kk) configuration.}
\end{figure}

To visualize and compare the overall hyperfine landscape presented by the different candidate structures discussed in the main text (kk, hh, kk+Csi(0,0,8), kk+Csi(0,0,8), OV(hh), OV(kk)), we analyze the statistical distribution of calculated hyperfine values $A_{zz} = |\hat{n}_0 \cdot \mathbf{A}^I\cdot \hat{n}_0|$ across all relevant $\mathrm{^{29}Si}$ and $\mathrm{^{13}C}$ lattice sites, where $\hat{n}_0$ is the unit vector pointing along the c axis. These results are shown in Fig.~\ref{hyper_stat}, Each panel facilitates the comparision of a specific candidate defect's hyperfine distribution against the baseline kk and hh divacancies, visually complementing the quantitative analysis of relative peak shifts presented in the main text.

\newpage
\ 
\newpage

\section{Other Potential Spin Defects}

\setlength{\LTcapwidth}{0.88\textwidth}
    \begin{longtable}{lccccccc}
      \caption{\label{tab:other_spin_defects}A broader survey of DV-antisite complexes investigated. While the analysis in the main paper now identifies the OV(hh) model as the most probable origin for the PL6 center, the DV-antisite complexes listed here were systematically screened based on their calculated zero-field splitting (ZFS) parameters D and E, the angle $\theta$ between the $D_{zz}$ principal axis and the lattice c-axis, Kohn-Sham (KS) gap, and binding energy ($E_b$). This table includes defects with negative binding energies, indicating potential stability. The term "red-shifted PL" indicates a ZPL substantially red-shifted relative to PL1-4. "Undetectable" labels defects with excessively large |E| values, typically not resolved in standard ESR/ODMR. Underlined KS gap values highlight the reason for "red-shifted PL" identification. Defects lacking sufficient constraints for definitive identification beyond these parameters are marked "unknown". The PL6 entries ($\mathrm{kk+C_{Si}(0,0,8)}$ and $\mathrm{kk+C_{Si}(0,0,\overline{8})}$) are included as they were the outcome of the DV-antisite screening detailed in the main text before comparision with the OV model.}\\

      \hline\hline
      defect & spin & $D$ (GHz) & $E$ (MHz) & $\theta$ & KS gap (eV) & $E_b$ (eV) & identification \\ 
      \hline
      \endfirsthead

      \caption[]{continued from previous page} \\
      \hline\hline
      defect & spin & $D$ (GHz) & $E$ (MHz) & $\theta$ & KS gap (eV) & $E_b$ (eV) & identification \\
      \hline
      \endhead

      \hline
      \multicolumn{8}{r}{{Continued on next page}} \\
      \endfoot

      \hline\hline
      \endlastfoot

      hh     & 2 &  1.506 &   6.4 & 0.1 & 1.118 & --     & PL1 \\
      kk     & 2 &  1.455 &   -4.3 & 0.2 & 1.141 & --     & PL2 \\
      kh     & 2 &  1.395 &  -61 & 109.9 & 1.111 & --     & PL3 \\
      hk     & 2 &  1.452 &  29 & 70.8 & 1.137 & --     & PL4 \\
      \hline
      $\mathrm{kk+  Si_C}(\overline{1}, \overline{1}, 1)$  & 0 &  0.000 &  0.0 & -- & 0.555 & -4.771 & -- \\
      $\mathrm{hh+  Si_C(\overline{1}, \overline{1}, 1)}$   & 0 &  0.000 &  0.0 & -- & 0.653 & -4.822 & -- \\[0.5ex]
      $\mathrm{kk+  C_{Si}}(0, \overline{2}, \overline{4})$  & 2 &  1.545 &  -26.9 & 0.0 & 1.137 & -0.964 & \textbf{unknown} \\
      $\mathrm{hh+  C_{Si}(\overline{1}, \overline{1}, \overline{4})}$  & 2 &  1.607 & -126.1 & 1.1 & 1.154 & -0.937 & \textbf{unknown}\\
      $\mathrm{kk+  Si_C}(2, 0, \overline{3})$    & 2 &  1.411 &  -23.3 & 177.5 & \underline{0.951} & -0.810 & red-shifted PL \\
      $\mathrm{hh+  Si_C(\overline{1}, \overline{3}, \overline{3})}$  & 2 &  1.470 & -109.9 & 4.3 & \underline{0.971} & -0.754 & red-shifted PL \\
      $\mathrm{kk+  Si_C}(3, \overline{1}, 1)$  & 2 &  1.287 & 144.7 & 0.8 & \underline{0.899} & -0.387 & red-shifted PL \\
      $\mathrm{hh+  Si_C(3, \overline{1}, 1)}$    & 2 &  1.338 & 158.9 & 1.6 & \underline{0.880} & -0.363 & red-shifted PL \\[0.5ex]
      $\mathrm{kk+  Si_C}(0, 0, 5)$  & 2 &  0.981 &   -6.9 & 0.2 & \underline{0.603} & -0.339 & red-shifted PL \\
      $\mathrm{kk+  C_{Si}}(2, 2, 0)$  & 2 &  1.306 & \underline{417.5} & 5.9 & 1.021 & -0.338 & undetectable \\
      $\mathrm{hh+  Si_C(\overline{2}, \overline{2}, 5)}$  & 2 &  1.049 &  94.5 & 1.6 & \underline{0.798} & -0.328 & red-shifted PL \\
      $\mathrm{kk+  C_{Si}}(\overline{1}, \overline{1}, 4)$  & 2 &  1.702 & \underline{-384.4} & 7.6 & 1.038 & -0.327 & undetectable \\
      $\mathrm{hh+  C_{Si}(\overline{1}, \overline{3}, 0)}$  & 2 & 1.347 &  \underline{459.9} & 7.0 & 1.001 & -0.321 & undetectable \\
      $\mathrm{hh+  Si_C(\overline{1}, \overline{1}, \overline{7})}$  & 2 &  1.425 & \underline{-439.8} & 11.2 & \underline{0.664} & -0.315 & undetectable \\
      $\mathrm{hh+  C_{Si}(\overline{1}, \overline{1}, 4)}$ & 2 & 1.769 & \underline{-433.2} & 8.5 & 1.034 & -0.309 & undetectable \\
       $\mathrm{kk+  Si_C}(4, 0, \overline{3})$   & 2 &  1.487 &  29.7 & 0.8 & 1.040 & -0.253 & \textbf{unknown} \\
        $\mathrm{kk+  Si_C}(0, \overline{2}, \overline{7})$  & 2 &  1.403 &  95.6 & 1.7 & \underline{0.900} & -0.241 & red-shifted PL \\
        $\mathrm{hh+  Si_C(4, 0, \overline{3})}$  & 2 &  1.544 &  -27.4 & 1.0 & 1.041 & -0.238 & \textbf{unknown} \\
      $\mathrm{kk+  C_{Si}(0, 0, \overline{8})}$  & 2 &  \underline{1.478} &  \underline{-4.3} & \underline{0.3} & \underline{1.176} & \underline{-0.204} & PL6 \\
      $\mathrm{hh+  C_{Si}(2,2,\overline{4})}$ & 2 & 1.523 & -36.7 & 0.6 & 1.138 & -0.183 & \textbf{unknown} \\
      $\mathrm{kk+  C_{Si}(\overline{2},\overline{2},\overline{4})}$ & 2 & \underline{1.460} & \underline{-43.8} & 0.2 & \underline{1.153} & -0.159 & hk-like \\
      $\mathrm{hh+  C_{Si}(0,\overline{2},\overline{8})}$ & 2 & \underline{1.527} & \underline{-14.8} & \underline{179.5} & \underline{1.111} & -0.098 & hh mixed \\
      $\mathrm{kk+  C_{Si}(4,0,0)}$ & 2 & 1.319 & -39.1 & 0.6 & 1.031 & -0.092 & \textbf{unknown} \\
      $\mathrm{kk+  C_{Si}(0, 0, 8)}$  & 2 & \underline{1.554} & \underline{0.3} & \underline{0.3} & \underline{1.207} & -0.083 & PL6 \\
      $\mathrm{hh+  C_{Si}(4, 0, 0)}$  & 2 &  1.387 &  -32.1 & 0.7 & 1.038 & -0.082 & \textbf{unknown} \\[0.5ex]
      $\mathrm{kk+  Si_C(2,0,5)}$ & 2 & 1.258 & 60.3 & 1.4 & \underline{0.839} & -0.066 & red-shifted PL \\
      $\mathrm{hh+  C_{Si}(\overline{2},\overline{2},8)}$ & 2 & 1.491 & -35.4 & 1.5 & 1.098 & -0.058 & \textbf{unknown} \\
      $\mathrm{hh+  Si_C(\overline{2},\overline{2},\overline{11})}$ & 2 & 1.486 & -44.5 & 0.6 & \underline{1.001} & -0.047 & red-shifted PL \\
      $\mathrm{hh+  Si_C(4,\overline{2},5)}$ & 2 & 1.484 & -23.3 & 0.2 & \underline{0.903} & -0.041 & red-shifted PL \\
      $\mathrm{kk+  C_{Si}(2,0,8)}$ & 2 & \underline{1.453} & \underline{46.1} & 1.3 & \underline{1.124} & -0.037 & hk-like \\
      $\mathrm{kk+  C_{Si}(3,3,0)}$ & 2 & \underline{1.477} & \underline{18.7} & 0.6 & \underline{1.151} & -0.035 & hk-like \\
      $\mathrm{kk+  Si_C(4,0,5)}$ & 2 & 1.458 & -36.8 & 0.1 & \underline{0.900} & -0.029 & red-shifted PL \\
      $\mathrm{kk+  C_{Si}(\overline{1},\overline{1},\overline{12})}$ & 2 & \underline{1.464} & \underline{1.0} & \underline{179.8} & \underline{1.149} & -0.024 & kk mixed \\
      $\mathrm{kk+  Si_C(3,\overline{3},5)}$ & 2 & 1.466 & -21.7 & 0.4 & \underline{0.885} & -0.024 & red-shifted PL \\
      $\mathrm{kk+  C_{Si}(2,0,\overline{8})}$ & 2 & \underline{1.469} & -16.3 & \underline{0.3} & \underline{1.158} & -0.020 & kk mixed \\
      $\mathrm{kk+  C_{Si}(0,\overline{2},12)}$ & 2 & \underline{1.473} & \underline{9.7} & \underline{0.1} & \underline{1.155} & -0.014 & kk mixed \\
      $\mathrm{hh+  C_{Si}(\overline{1},\overline{1},\overline{12})}$ & 2 & \underline{1.518} & \underline{-10.0} & \underline{0.1} & \underline{1.127} & -0.014 & hh mixed \\
      $\mathrm{hh+  Si_C(3,\overline{1},\overline{7})}$ & 2 & \underline{1.506} & \underline{-9.6} & \underline{0.6} & 1.072 & -0.013 & hh mixed \\
      $\mathrm{hh+  C_{Si}(0, \overline{2}, 8)}$  & 2 &  \underline{1.559} &  \underline{-13.2} & \underline{0.6} & \underline{1.109} & -0.010 & hh mixed \\
  \end{longtable}

  The complexes listed in Table~\ref{tab:other_spin_defects} represent a subset of the DV-antisite configurations screened, fousing on those with binding energies at or below -0.01~eV, suggesting potential stability (formation dynamics were not considered). While our main analysis now points to the OV(hh) model as the most probable origin for the PL6 center, this broader dataset of DV-antisite complexes, including the $\mathrm{kk+C_{Si}(0,0,8)}$ and $\mathrm{kk+C_{Si}(0,0,\overline{8})}$ configurations that initially emerged as strong DV-antisite candidates for PL6, may harbor other, yet unidentified, spin-active centers in SiC.
  
  Several general observations can be made from this DV-antisite survey. The properties of divacancies primarily originate from the C dangling bonds formed by the $V_\mathrm{Si}$ within the divacancy. Introducing a nearest-neighbor antisite ${\rm Si_C}$ disrupts these C dangling bonds, resulting in an $S=0$ ground state, as observed in the $\mathrm{kk+  Si_C}(\overline{1}, \overline{1}, 1)$ and $\mathrm{hh+  Si_C(\overline{1}, \overline{1}, 1)}$ configurations. Electron pairing in these defects enhances their stability relative to others. Furthermore, most other ${\rm Si_C}$-perturbed complexes exhibit significantly reduced crystal field splitting, result in lower-energy PL (typically less than 1.0~eV) or complete PL quenching, rendering them absent in PL spectra. Coversely, some complexes display very large $|E|$ values (e.g. 300-400~MHz for certain configurations), whose resonances are too widely separated in ESR or ODMR spectra for unambiguous identification. Moreover, for distant antisites, weak interactions with the central divacancy result in small perturbation, making them hard to distinguish from their parent divacancies. Notably, many defects labeled as ``kh-like'' exhibit spin properties similar to the hk divacancy while retaining ZPL energies near their original values. However, the principal axis of the $D_{zz}$ remains approximately aligned with the c-axis, a feature resolvable through angle-resolved EPR experiments. 

Among the numerous stable defect complexes identified in our DV-antisite screening, a limited subset exhibits characteristics that warrant investigation as potentially distinct spin centers. While the $\mathrm{kk+ C_{Si}(0, 0, \overline{8})}$ and $\mathrm{kk+C_{Si}(0, 0, 8)}$ configurations were initially considered strong candidates for PL6 within this framework, our broader analysis (detailed in the main paper) now favors an OV(hh) origin for PL6. Nevertheless, these two DV-antisite complexes, along with eight other stable configurations highlighted in bold within Table~\ref{tab:other_spin_defects}, present intriguing properties that suggest they could be effective, novel spin centers in their own right. Notably, the $\mathrm{hh+Si_C(4, 0, \overline{3})}$ and $\mathrm{kk+Si_C(4, 0, \overline{3})}$ exhibit similar structural compositions and comparable properties. Although their KS level gaps are narrower than those of PL1-4, resulting in lower ZPL energies, their $D$ and $E$ parameters remain comparable to those of PL1-4. A distinct pair, kk+$\mathrm{C_{Si}(4,0,0)}$ and hh+$\mathrm{C_{Si}(4,0,0)}$, demonstrates reduced ZPL energies alongside significantly smaller $D$ values. The most stable spin defect complexes, kk+$\mathrm{C_{Si}(0, \overline{2}, \overline{4})}$ and hh+$\mathrm{C_{Si}(\overline{1}, \overline{1}, \overline{4})}$, exhibit ZPL energies within the PL1-4 spectral range. While kk+$\mathrm{C_{Si}(0, \overline{2}, \overline{4})}$ shares similarities with the hk defect, it features an enhanced $D$ parameter. The hh+$\mathrm{C_{Si}(\overline{1}, \overline{1}, \overline{4})}$ complex displays an exceptionally large $E$ value and a $D$ parameter exceeding those of PL1-4 while simultaneously demonstrating superior thermodynamic stability based on binding energy analysis. Other notable configurations include hh+$\mathrm{C_{Si}(2,2,\overline{4})}$, which exhibits elevated ZPL energy, increased $|E|$ magnitude, and a higher $D$ value relative to the hh defect, and hh+$\mathrm{C_{Si}(\overline{2},\overline{2},8)}$, which shows reduced ZPL energy while retaining hk-like defect characteristics.

\nocite{*}
\bibliographystyle{apsrev4-2}
\bibliography{supplement}